\documentclass[twocolumn,showpacs,preprintnumbers,amsmath,amssymb,prb]{revtex4}

\usepackage{graphicx}
\usepackage{dcolumn}
\usepackage{bm}
\usepackage{color}

\begin{document}


\title{Quantum phases of hardcore bosons with long-range interactions on a square lattice
}

\author{Daisuke Yamamoto$^{1}$}
\email{d-yamamoto@riken.jp}
\author{Akiko Masaki$^{2}$}
\author{Ippei Danshita$^{3}$}
\affiliation{
{$^1$Condensed Matter Theory Laboratory, RIKEN, Wako, Saitama 351-0198, Japan}
\\
{$^2$Institute for Solid State Physics, University of Tokyo, Kashiwa, Chiba 277-8581, Japan}
\\
{$^3$Computational Condensed Matter Physics Laboratory, RIKEN, Wako, Saitama 351-0198, Japan}
}
\date{\today}

\begin{abstract}
We study the ground-state phase diagrams of hardcore bosons with long-range interactions on a square lattice using the linear spin-wave theory and a cluster mean-field method. Specifically, we consider the two types of long-range interaction: One consists only of the nearest- and next-nearest-neighbor interactions, and the other is the dipole-dipole interaction that decays with the interparticle distance $r$ as $\sim r^{-3}$. It is known from previous analyses by quantum Monte Carlo methods that a checkerboard supersolid (CSS) is absent in the ground-state phase diagram of the former case while it is present in the latter. In the former, we find that quantum fluctuations around mean-field solutions are enhanced by the direct competition between the checkerboard and striped solid orders and that they destabilize the CSS phase. On the other hand, the emergence of the CSS phase in the latter case can be attributed to the absence of such a competition with other solid orders. We also show that the cluster mean-field method allows for the determination of phase boundaries in a precise quantitative manner when scaling with respect to the cluster size is taken into account. It is found that the phase transition between the superfluid and the solid (or CSS) is of the first order in the vicinity of the particle-hole symmetric line. 
\end{abstract}

\pacs{03.75.-b, 05.30.Jp, 67.80.kb}
\maketitle
\section{\label{1}Introduction} 
Can a solid exhibit superfluidity in lattice systems? 
This question was first investigated theoretically by Matsuda and Tsuneto~\cite{matsuda-70,liu-72} in the context of the quantum lattice-gas model for $^4$He, which assumes that atoms move only on fixed lattice points even in the liquid phase.~\cite{matsubara-56} 
Using the lattice representation, they discussed the possibility of supersolidity, which is characterized by the coexistence of solid (diagonal) and superfluid (off-diagonal) long-range orders, in bulk and thin film of $^4$He. In the lattice system, the continuous translational invariance of the system is broken by the presence of the background discrete structure, and the ``solid'' means a state in which a discrete translational invariance is broken spontaneously. Recently, this issue has attracted renewed interest in connection with ultracold Bose gases in optical lattices. The creation of gases with strong dipole-dipole interactions~\cite{griesmaier-05, lu-11, aikawa-12, ni-08, aikawa-10,deiglmayr-08} has provided an ingredient essential for the emergence of supersolid phases, namely long-range interactions. Moreover, the precise controllability of optical-lattice systems has inspired theoretical explorations of supersolid phases in various types of lattice structure, such as chain,~\cite{batrouni-06, burnell-09} square,~\cite{goral-02,kovrizhin-05, sengupta-05,scarola-05,yi-07,danshita-09,sansone-10,danshita-10} triangular,~\cite{wessel-05,heidarian-05,melko-05,boninsegni-05,hassan-07,sen-08,pollet-10,bonnes-11,zhang-11,yamamoto-12} honeycomb,~\cite{wessel-07,gan-07} kagome,~\cite{isakov-06} and cubic~\cite{yi-07,kyamamoto-09,xi-11,ohgoe-12} lattices. We also note that the formation of checkerboard density-wave order has been experimentally observed in Bose-Einstein condensates coupled with an optical cavity.~\cite{baumann-10}

For understanding lattice supersolids, it is important to address the following questions:
in what situations the coexistent state can emerge and why it can be stable in such situations. 
Extensive studies over the past few decades have provided answers to these questions.
For example, previous researches demonstrated that no supersolid phases can exist in the ground-state phase diagram of the hardcore Bose-Hubbard model with the nearest-neighbor (NN) interaction for bipartite lattices such as square~\cite{batrouni-00} and honeycomb~\cite{wessel-07, gan-07} lattices. 
In these cases, uniform supersolid states are unstable towards the formation of domain walls,~\cite{zhang-03,sengupta-05} and the system undergoes phase separation into superfluid and solid phases. 
In order for supersolid phases to be present,
one has to modify the model by, e.g., introducing dipole-dipole interactions~\cite{sansone-10,ohgoe-11} or treating softcore bosons.~\cite{kovrizhin-05,sengupta-05} 
In contrast, the triangular-lattice system of hardcore bosons with only the NN interaction has stable supersolid phases for the fillings $1/3<\rho <2/3$.~\cite{murthy-97,wessel-05,bonnes-11,zhang-11,yamamoto-12} 
As for the case of the kagome lattice, although the mean-field (MF) analysis predicts the existence of supersolid states,~\cite{murthy-97} they are destabilized by the effects of strong quantum fluctuations.~\cite{isakov-06}

In this paper, focusing on the supersolid phase with checkerboard solid order, we analyze ground-state properties of hardcore bosons with long-range interactions on a square lattice by means of the linear spin-wave (LSW) theory and a cluster mean-field (CMF) method. In this system, 
the range of the interactions makes a qualitative difference in the emergence of checkerboard supersolid (CSS) states. 
The previous quantum Monte Carlo (QMC) calculations~\cite{batrouni-00} have shown that no CSS phase is present between the superfluid (SF) and checkerboard solid (CS) phases in the system with only the NN interaction $V_1$ and the next-nearest-neighbor (NNN) interaction $V_2$.
On the other hand, it is known that the infinite-range dipole-dipole interaction, which decays as the inverse cube of the distance, can stabilize the CSS states.~\cite{sansone-10}
We will clarify the reasons why the dipole-dipole interaction can stabilize the CSS states unlike the case of only the NN and NNN interactions.

The MF ground-state (classical) properties of the hardcore Bose-Hubbard models with dipole-dipole interaction and with only the NN and NNN interactions have already been discussed separately in previous works.~\cite{danshita-10,bruder-93,scalettar-95,pich-98} We first review those results from the standpoint of comparing the two types of interactions. When assuming that the system is in the phases with checkerboard (two-sublattice) order, the MF energy of the dipolar model can be naturally written in the same form as that of the model with effective NN and NNN interactions, $V_1^{\rm eff}$ and $V_2^{\rm eff}$. We find that the value of $V_2^{\rm eff}/V_1^{\rm eff}$ is very large, and it leads to a large region of CSS phase in the ground-state phase diagram at the MF level.  
Second, from the LSW analysis, we show that quantum fluctuations around the MF solutions are not so strong compared to the case of only the NN and NNN interactions, which is attributed to the absence of the direct competition between the checkerboard and other solid orders. These two factors lead to the emergence of the stable CSS state in the dipolar system unlike the case of the shorter-range interactions.

Moreover, including the effects of quantum fluctuations, we derive the ground-state phase diagrams. Although some of the results have already been known from previous QMC works, we reconsider the issue in detail in terms of another numerical approach based on a large-size CMF method.~\cite{yamamoto-12} From a comparison with the QMC result~\cite{sansone-10} for the model with dipole-dipole interaction, it is shown that the CMF method combined with cluster-size scaling can locate the phase boundaries quantitatively.
We also derive the phase diagram of the model with only the NN and NNN interactions and confirm that the region of stable CSS phase almost completely disappears due to the strong quantum fluctuations.
Moreover, we find the first-order phase transition between the SF and the CS (or the CSS) in the close vicinity of the particle-hole symmetry line for the both models. It is worth stressing that 
our CMF procedure is free from the minus-sign problem even when applying to frustrated systems. Moreover, it is useful to study metastability phenomena such as hysteresis,~\cite{yamamoto-12} since one can get all stationary points of the free energy including metastable and saddle-point solutions.

The remainder of the paper is organized as follows. In Sec.~\ref{2}, we introduce our models describing hardcore bosons with 
two types of long-range interactions in a square lattice. 
In Sec.~\ref{3}, we show the ground-state phase diagrams of the two models within the mean-field theory. 
In Sec.~\ref{4}, we perform the LSW analyses to discuss the strength of quantum fluctuations around the MF solutions. 
In Sec.~\ref{5}, applying a CMF method and the cluster-size scaling, we obtain the phase diagrams including the effects of the quantum fluctuations. Moreover, we summarize the reasons why the dipole-dipole interaction stabilizes the CSS states, based on the results obtained in Secs.~\ref{3}-\ref{5}. 
The conclusion is given in Sec.~\ref{6}.

\section{\label{2} Hardcore Bose-Hubbard Models}
We consider interacting hardcore bosons on a square lattice given by the following Hamiltonian:
\begin{eqnarray}
\hat{H}=
-J\sum_{\langle j,l \rangle}
(\hat{a}^{\dagger}_{j} \hat{a}_{l}+{\rm H.c.})
+\frac{1}{2}\sum_{ j, l }
V_{jl} \hat{n}_{j}\hat{n}_{l}
-\mu \sum_j \hat{n}_j,
\label{hamiltonian}
\end{eqnarray}
where $\hat{a}^{\dagger}_{j}$ and $\hat{n}_j=\hat{a}^{\dagger}_{j} \hat{a}_{j}$ are the creation and number operators of the hardcore bosons at site $j$, $J$ denotes the hopping amplitude between NN pairs, and $\mu$ the chemical potential. 
The hardcore boson limit means the situation where two or more bosons are not allowed to occupy the same site due to the strong on-site interaction $U\rightarrow\infty$. We assume the existence of a long-range interaction $V_{ij}$ between the hardcore bosons and consider two different forms of $V_{jl}$ such that we study the effect of long-range interactions on the stability of supersolid phases through the comparison of the two models. 

\begin{figure}[t]
\includegraphics[scale=0.65]{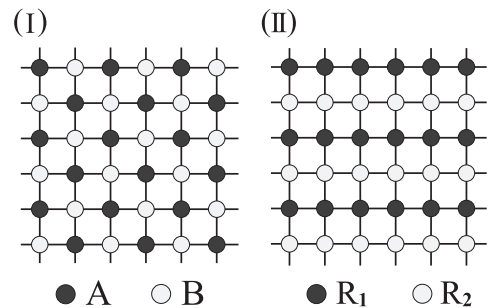}
\caption{\label{sublattice}
Schematic pictures of the sublattice structure of the (I) checkerboard and (II) stripe patterns. The circles indicate the sites of the square lattice and the lattice sites of the same color belong to the same sublattice. }
\end{figure}

The first one is given by
\begin{eqnarray}
V_{jl}&=&\left\{ \begin{array}{ll}
V_1 & (\left|{\bf r}_{j}-{\bf r}_{l}\right|=d) \\
V_2 & (\left|{\bf r}_{j}-{\bf r}_{l}\right|=\sqrt{2}d) \\
0 & ({\rm otherwise}) ~~~~~~~~~~~[V_1{\rm -}V_2~{\rm model}].
\end{array} \right.\label{V1V2}
\end{eqnarray}
where $d$ is the lattice spacing and ${\bf r}_j=(j_x d,j_y d)$ with integers $j_x$ and $j_y$ is a lattice vector at site $j$. The parameters $V_1\geq 0$ and $V_2\geq 0$ represent the strength of the NN and NNN interactions, respectively. The NN interaction $V_1$ tends to induce the checkerboard density-wave order depicted in Fig.~\ref{sublattice}(I), while the strong NNN interaction $V_2$ favors the stripe pattern in Fig.~\ref{sublattice}(II).~\cite{pich-98,batrouni-00} 
Thus, the Hamiltonian in Eq.~(\ref{hamiltonian}) with Eq.~(\ref{V1V2}), which we refer to as the ``$V_1$-$V_2$ model,'' is a minimal model for studying the competition among two different solid orders and superfluidity induced by the hopping $J$.~\cite{bruder-93,batrouni-95,scalettar-95,pich-98,batrouni-00,ng-0810} We will focus on the regime of checkerboard ordering, $V_2/V_1\lesssim 1/2$,~\cite{batrouni-95,scalettar-95,pich-98}.

The second one is the isotropic dipole-dipole interaction that is more realistic from an experimental point of view. In experiments of ultracold gases, one of the most promising way to prepare long-range interacting systems is the use of the so-cold ``dipolar'' atoms, such as chromium,~\cite{griesmaier-05} dysprosium,~\cite{lu-11} and erbium,~\cite{aikawa-12} or molecules, such as KRb~\cite{ni-08,aikawa-10} and LiCs.~\cite{deiglmayr-08} These atoms and molecules have a large (magnetic or electric) dipole moment, which leads to strong long-range forces among the dipolar particles. We assume that the dipole moments are fully polarized along the direction perpendicular to the lattice plane. In this case, the interaction between the dipoles works isotropically and its long-range part can be well approximated by
\begin{eqnarray}
V_{jl}&=&\left\{ \begin{array}{ll}
Vd^3/\left|{\bf r}_{j}-{\bf r}_{l}\right|^3   &  (j\neq l) \\
0 & (j=l)
\end{array} \right.~~[V_{\rm dip}~{\rm model}].\label{Dip}
\end{eqnarray}
We refer to the model given by the Hamiltonian in Eq.~(\ref{hamiltonian}) with Eq.~(\ref{Dip}) as the ``$V_{\rm dip}$ model,'' hereafter. 
The dipole-dipole interaction falls off as the inverse cube of the distance as $\{V,0.354V,0.125V,0.089V,\cdots\}$ (for the NN, NNN, third, fourth neighbors). Therefore, it appears that most of the essential physics can be captured with just the first two terms, namely within the $V_1$-$V_2$ model. In fact, as will be shown in the next section, the MF phase diagrams of the two models are very similar; there are regions of the standard SF phase, solid phases, and supersolid phases, including the CSS phase.

However, previous numerical analyses based on the QMC method demonstrated that the correct ground-state phase diagrams, which include quantum fluctuations, have a crucial difference between the finite-range $V_1$-$V_2$ model and the infinite-range $V_{\rm dip}$ model. For the $V_1$-$V_2$ model, the authors of Ref.~\onlinecite{batrouni-00} concluded that the CSS phase predicted by the MF theory is completely destabilized by strong quantum fluctuations and does not appear in the QMC calculations for any value of $V_2/V_1$, although they checked it only for $V_1=3J$. In contrast, as for the $V_{\rm dip}$ model, the main features of the MF phase diagram can survive,~\cite{sansone-10} including the existence of the CSS phase. This indicates that the long-range part of the dipole-dipole interactions plays a crucial role in stabilizing the CSS phase. 
In the following sections, we shall analyze the ground states of the two models and discuss the role of the long-range interactions in the emergence of the CSS state in order to clarify the reasons why the two models have the qualitative difference.

The instability of supersolid phases against phase separation has often been discussed from a perturbative point of view assuming that $J/V_{jl}$ is small; the total energy gains from the lowest-order hopping process of doped bosons (or holes) and from the surface energy are compared on classical solid states with/without a domain-wall.~\cite{zhang-03,sengupta-05,wessel-05,ohgoe-12} 
However, in the $V_{\rm dip}$ model, the CSS states can appear even for relatively large values of $J/V_{jl}$, and the structure called the $devil's$ $staircase$ with many different types of solid states emerges in the region of small values of $J/V_{jl}$.~\cite{sansone-10} Hence, we will present a more careful discussion from a different angle by using the LSW and CMF methods.

\section{\label{3}Classical Ground States}
To begin with, we show the ground-state properties within the MF theory. From the equivalence of the hardcore-boson and spin-1/2 operators,~\cite{matsubara-56} Eq.~(\ref{hamiltonian}) can be mapped onto the spin-$1/2$ $XXZ$ model with long-range Ising-type interactions: 
\begin{eqnarray}
\hat{H}_{\rm spin}&=&
-2J\sum_{\langle j,l \rangle}
\left(\hat{S}^{x}_{j} \hat{S}^{x}_{l}+\hat{S}^{y}_{j} \hat{S}^{y}_{l}\right)+\frac{1}{2}\sum_{ j , l }V_{jl} \hat{S}^{z}_{j}\hat{S}^{z}_{l}\nonumber\\
&&
-h \sum_j \hat{S}^{z}_j,
\label{spinH}
\end{eqnarray}
where $\hat{\bf S}_j=(\hat{S}^x_j,\hat{S}^y_j,\hat{S}^z_j)$ is the pseudospin operator which satisfies the commutation relations
\begin{eqnarray}
[\hat{S}_j^\mu,\hat{S}_l^\nu]=i\epsilon_{\mu\nu\lambda} \hat{S}_j^\lambda\delta_{jl}.\label{rule}
\end{eqnarray}

In the pseudospin language, the occupied and unoccupied states of bosons correspond to the spin-up and spin-down states, respectively. Thus the filling factor, which is the average density per site, of hardcore bosons can be calculated through the relation $\rho\equiv\sum_j\langle\hat{n}_j\rangle/M= 1/2+\sum_j\langle\hat{S}_j^z\rangle/M$. Here, $M$ is the number of lattice sites. 
The pseudospin raising and lowering operators $\hat{S}_j^\pm=\hat{S}^x_j+i\hat{S}^y_j$ play the role of the creation and annihilation of the hardcore bosons; $\hat{a}_j^\dagger =\hat{S}_{j}^{+},~~\hat{a}_j = \hat{S}_{j}^{-}$. 
The effective magnetic field acting on the pseudospins is given by $h=\mu -z\bar{V}/2$ with
\begin{eqnarray}
\bar{V} &\equiv& \frac{1}{z}\sum_{j}V_{0j}\nonumber \\
&=&\left\{ \begin{array}{ll}
V_1+V_2   &  ({\rm for}~{\rm the}~V_1{\rm -}V_2~{\rm model}) \\
 2.258V & ({\rm for}~{\rm the}~V_{\rm dip}~{\rm model})
\end{array} \right.,
\end{eqnarray}
where $z=4$ is the coordination number of the square lattice.
As can be obviously seen from the definition, the zero magnetic field corresponds to the particle-hole symmetric point ($\mu=z\bar{V}/2$) of the hardcore-boson model. 
Moreover, the density $\rho_j$ and the condensate wave function $\Psi_j$ of bosons are expressed by the longitudinal components $\langle \hat{S}^z_j\rangle$ and the transverse components $\langle \hat{S}^-_j\rangle$ as $\rho_j = \langle \hat{S}^z_j\rangle +1/2$ and $\Psi_j = \langle \hat{S}^-_j\rangle$.~\cite{matsuda-70,liu-72} 
From these correspondences, we can use the calculation methods which have been developed in the field of quantum spins for studying hardcore-boson systems.

At zero temperature, replacing the local pseudospin operators in Eq.~(\ref{spinH}) with the classical vectors of length $S=1/2$, 
\begin{eqnarray}
\hat{\bf S}_j\rightarrow {\bf S}_j^{\rm cl}=S( \cos \varphi_j \sin \theta_j,\sin \varphi_j \sin \theta_j,\cos \theta_j), 
\end{eqnarray}
we obtain the MF (classical) energy as a function of the orientation of the local pseudospins $\{\theta_j,\varphi_j\}$:
\begin{eqnarray}
E_{0}&=&-\frac{S^2}{2}\sum_{j,l}\Big[2J_{jl}\sin \theta_j\sin \theta_l \cos (\varphi_j-\varphi_l)\nonumber\\
&&-V_{jl}\cos \theta_j\cos \theta_l \Big]-hS\sum_j\cos \theta_j, \label{eq:EneCl}
\end{eqnarray}
where
\begin{eqnarray}
J_{jl}&=&\left\{ \begin{array}{ll}
J & ~~~~(|{\bf r}_j-{\bf r}_l|=d) \\
0 & ~~~~({\rm otherwise}) 
\end{array} \right..
\end{eqnarray}
Minimizing the MF energy with respect to $\{\theta_j,\varphi_j\}$, we derive the classical pseudospin configurations in the usual manner,~\cite{bruder-93,scalettar-95,pich-98} and translate the results into the hardcore-boson language. Without loss of generality, we can take $\varphi_j=0$, which means that the canted spins are assumed to lie in the $xz$ plane. The procedure described here gives the same results as those obtained by the standard decoupling technique for the intersite spin-exchange interaction terms (i.e., the Weiss molecular-field theory) at $T=0$. 

\subsection{\label{3-1}The MF results for the $V_1$-$V_2$ model}
\begin{figure*}[t]
\includegraphics[scale=0.8]{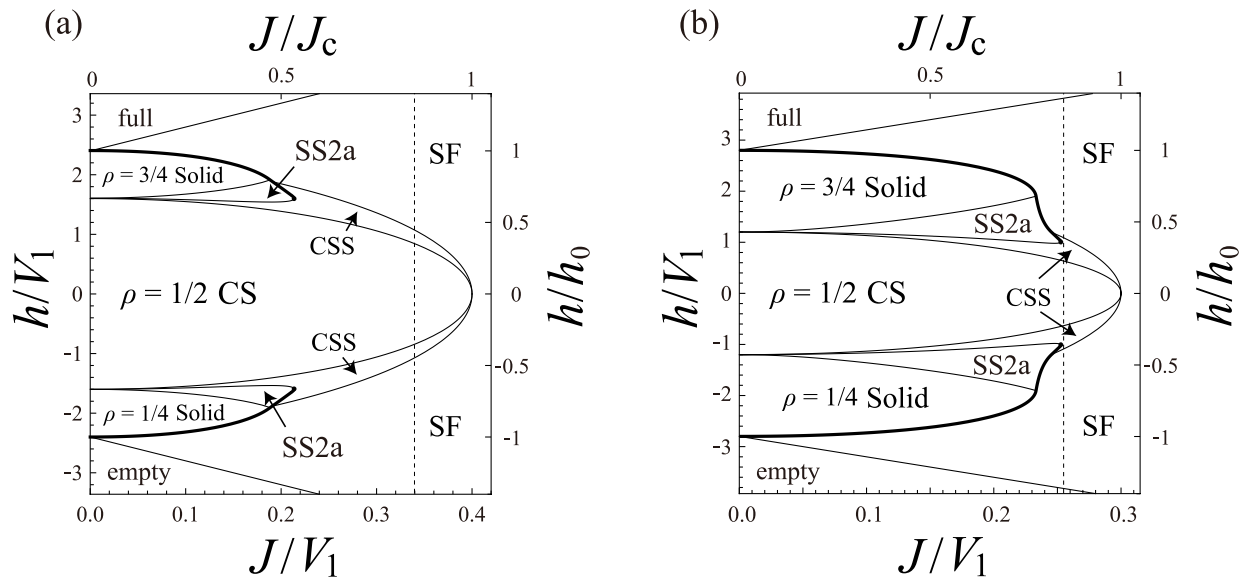}
\caption{\label{MFPDV1V2}
Ground-state MF phase diagrams of the $V_1$-$V_2$ model in the $(J,h)$ plane for (a) $V_2/V_1=0.2$ (after Pich $et$ $al$., Ref.~\onlinecite{pich-98}) and (b) $V_2/V_1=0.4$. Second- and first-order phase transitions are indicated by the thin and thick solid lines, respectively. The dashed vertical lines mark the location of $J/J_{\rm c}=0.85$. The axes are scaled in two different ways: by $V_1$ for both axes and by $J_{\rm c}$ and $h_0$, respectively. 
}
\end{figure*}
In this subsection, let us briefly review the MF results for the $V_1$-$V_2$ model.~\cite{bruder-93,scalettar-95,pich-98} We mainly focus on the phases with the two-sublattice structure described in Fig.~\ref{sublattice}(I). Within this checkerboard structure, we can describe the CS and CSS states in addition to the uniform SF state. The CS state, which is an insulating state appearing at half filling, has the checkerboard density-wave order characterized by $\rho_{\bf Q}\equiv \sum_j \langle \hat{n}_j\rangle\exp (i{\bf Q}\cdot {\bf r}_j) /M$ with ${\bf Q}=(\pi/d,\pi/d)$, while the SF state has the off-diagonal long-range order characterized by $\Psi\equiv \sum_j \langle \hat{a}_j\rangle/M$. The CSS state has both of the two (diagonal and off-diagonal) orders. 
In addition, completely empty ($\rho=0$) and fully occupied ($\rho=1$) states also appear. These trivial incompressible states can be regarded as a kind of Mott insulator (MI) states.

\begin{figure}[b]
\includegraphics[scale=0.6]{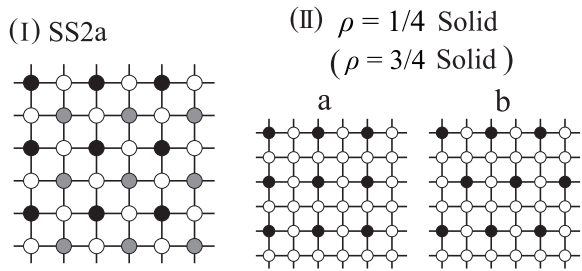}
\caption{\label{sublatticesV1V2}
Schematic pictures of the (I) SS2a and (II) $\rho=1/4$ ($\rho=3/4$) solid phases emerging in the phase diagram of the $V_1$-$V_2$ model. There are two possibilities (a and b) for the structure of the $\rho=1/4$ ($\rho=3/4$) solid phase, which are energetically degenerate for the $V_1$-$V_2$ model within the MF theory. 
}
\end{figure}

In the classical limit, these states can be expressed in terms of pseudospin angles as follows:
\begin{eqnarray}
\begin{array}{ll}
\cos \theta_{\rm A}  = - \cos \theta_{\rm B}  = 1&  ({\rm CS})  \\
\theta_{\rm A} \neq \theta_{\rm B}~{\rm and}~\sin \theta_{\rm A},
\sin \theta_{\rm B}  \neq 0& ({\rm CSS})  \\
\sin \theta_{\rm A} =\sin \theta_{\rm B}  \neq 0& ({\rm SF})  \\
\cos \theta_{\rm A} =\cos \theta_{\rm B}  = 1~{\rm or}~-1& ({\rm MI})
\end{array},
\end{eqnarray}
where $\theta_{\rm A} $ and $\theta_{\rm B} $ are the canting angles of the pseudospins on sublattices ${\rm A} $ and ${\rm B} $ [see Fig.~\ref{sublattice}(I)]. The MF energy in Eq.~(\ref{eq:EneCl}) per site can be rewritten as a function of $\theta_{\rm A} $ and $\theta_{\rm B} $:
\begin{eqnarray}
E_{0}^{(\rm ch)}/M&=&-4J S^2\sin \theta_{\rm A} \sin \theta_{\rm B} +2V_1S^2\cos \theta_{\rm A} \cos \theta_{\rm B} \nonumber\\&&+V_2S^2(\cos^2 \theta_{\rm A} +\cos^2 \theta_{\rm B} )\nonumber\\&&-hS(\cos \theta_{\rm A} +\cos \theta_{\rm B} )/2,
\label{EneCh}
\end{eqnarray}
and the filling factor is given by $\rho=1/2+S(\cos \theta_{\rm A}  + \cos \theta_{\rm B})/2$.
The ground-state phases are determined so as to minimize the MF energy $E_{0}^{(\rm ch)}$ with respect to $\theta_{\rm A} $ and $\theta_{\rm B} $. 
Figures~\ref{MFPDV1V2}(a) and \ref{MFPDV1V2}(b) show the phase diagrams in the ($J/V_1$, $h/V_1$)-plane for two different values of $V_2/V_1$. The value of $J$ at the tip of the CS phase, $J_{\rm c}$, is given by $(V_1-V_2)/2$ within the MF theory. The phase boundaries between CS and CSS, between CSS and SF, and between SF and MI are given by $h=\pm h_{{\rm c}1}$, $\pm h_{{\rm c}2}$, and $\pm h_{{\rm c}3}$, where
\begin{subequations}
\begin{eqnarray}
h_{{\rm c}1}&=&4S\sqrt{(V_1-V_2+2J)(V_1-V_2-2J)},\\
h_{{\rm c}2}&=&4S(V_1+V_2+2J)\sqrt{\frac{V_1-V_2-2J}{V_1-V_2+2J}},\\
h_{{\rm c}3}&=&4S(V_1+V_2+2J).
\end{eqnarray}\label{hcV1V2}\end{subequations}
In the figures, the quantities on the axes are also scaled by $J_{\rm c}$ and $h_0\equiv h_{{\rm c}3}|_{J=0}$ to compare the results in the same scale. In addition to the CS and CSS phases, other solid (with $\rho=1/4$, and $3/4$) and supersolid (SS2a) phases are formed due to the competition of the NN and NNN repulsions.~\cite{bruder-93,pich-98} These phases have the sublattice structures depicted in Fig.~\ref{sublatticesV1V2}. 

As seen in Eqs.~(\ref{hcV1V2}), the CSS phase can emerge as long as the NNN interaction is finite, and the window $h_{{\rm c}1}<|h|<h_{{\rm c}2}$ gets wider as $V_2$ increases. Thus, it appears that we just have to prepare the system with a stronger NNN interaction in order to obtain the stable CSS phase in a wider range of the parameters. 
However, when the value of $V_2/V_1$ is large,  the striped solid order shown in Fig.~\ref{sublattice}(II) is more favored than the checkerboard. The general expression of the MF energy for striped phases is given by
\begin{eqnarray}
E_{0}^{(\rm st)}/M&=&-JS^2 \left(\sin \theta_{\rm R_1}+\sin \theta_{\rm R_2}\right)^2\nonumber\\
&&+V_1S^2 \left(\cos \theta_{\rm R_1}+\cos \theta_{\rm R_2}\right)^2/2\nonumber\\
&&+2V_2S^2\cos \theta_{\rm R_1}\cos \theta_{\rm R_2}\nonumber\\
&&-hS(\cos \theta_{\rm R_1}+\cos \theta_{\rm R_2})/2,
\label{EneSt}
\end{eqnarray}
where $\theta_{\rm R_1}$ and $\theta_{\rm R_2}$ are the canting angles of the pseudospins on even and odd rows [see Fig.~\ref{sublattice}(II)]. For example, let us consider the solid orders emerging at the half-filling ($\rho=1/2$). Putting $\cos \theta_{\rm A}=-\cos \theta_{\rm B}=1$ in Eq.~(\ref{EneCh}) and $\cos \theta_{\rm R_1}=-\cos \theta_{\rm R_2}=1$ in Eq.~(\ref{EneSt}), we obtain the MF energies of the CS and striped solid states: 
\begin{subequations}
\begin{eqnarray}
E_{0}^{(\rm ch)}/M&=&-2S^2(V_1-V_2)~~ ({\rm CS}) ,\\
E_{0}^{(\rm st)}/M&=&-2S^2V_2~~ ({\rm striped}~{\rm solid}) . 
\end{eqnarray}
\end{subequations}
From the comparison, one finds that the striped solid state has lower energy than the CS state when $V_2/V_1>1/2$. Also for the supersolid phase, the striped one takes the place of the CSS phase in this regime.~\cite{pich-98} Because of the transitions to the striped phases, we cannot extend the CSS region by exceeding the limit of $V_2/V_1=1/2$. 
Moreover, as shown in Fig.~\ref{MFPDV1V2}(b), when the value of $V_2/V_1$ approaches the boundary to the stripe regime, the SS2a phase is extended toward the large $J/V_1$ region due to the competition of the two density-wave orders. This competition also causes strong quantum fluctuations that destabilize the CSS states, as will be discussed in Sec.~\ref{4}.

\subsection{\label{3-2}The MF results for the $V_{\rm dip}$ model}
Next, let us move onto the $V_{\rm dip}$ model. 
In Ref.~\onlinecite{danshita-09}, we have applied the MF theory to this model, and examined the stability of superflow in the CSS state. 
Here, we present more detailed information on the MF ground states, and discuss the comparison with the results for the $V_1$-$V_2$ model.

The MF energy per site of the $V_{\rm dip}$ model for the checkerboard pattern can be written as
\begin{eqnarray}
E_{0}^{(\rm ch)}/M&=&-4J S^2\sin \theta_{\rm A} \sin \theta_{\rm B} +2V_1^{\rm eff}S^2\cos \theta_{\rm A} \cos \theta_{\rm B} \nonumber\\&&+V_2^{\rm eff}S^2(\cos^2 \theta_{\rm A} +\cos^2 \theta_{\rm B} )\nonumber\\&&-hS(\cos \theta_{\rm A} +\cos \theta_{\rm B} )/2.
\label{EneChDip}
\end{eqnarray}
Here, $V_1^{\rm eff}$ ($V_2^{\rm eff}$) is just the summation of the long-range interactions between the pseudospins on the same (different) sublattice sites:
\begin{subequations}
\begin{eqnarray}
V_1^{\rm eff}&\equiv& \frac{1}{z}\sum_{l_{\rm B} } V_{j_{\rm A} l_{\rm B} }=1.460V,\\
V_2^{\rm eff}&\equiv& \frac{1}{z}\sum_{l_{\rm A} } V_{j_{\rm A} l_{\rm A} }= 0.7985V.
\end{eqnarray}\label{VeffCH}\end{subequations}
The index $j_{\rm A} $ ($j_{\rm B} $) means the $j$th site on sublattice ${\rm A} $ (${\rm B} $). 
Only by replacing $V_1$ and $V_2$ with $V_1^{\rm eff}$ and $V_2^{\rm eff}$ in Eq.~(\ref{EneCh}), we can immediately obtain the expression of Eq.~(\ref{EneChDip}). This means that the MF properties of the checkerboard phases of the $V_{\rm dip}$ model can be described exactly by the $V_1$-$V_2$ model with the effective NN and NNN interactions $V_1^{\rm eff}$ and $V_2^{\rm eff}$. For example, the phase boundaries between the CS, CSS, SF, and MI phases are obtained by replacing $V_1$ and $V_2$ in Eqs.~(\ref{hcV1V2}) with $V_1^{\rm eff}$ and $V_2^{\rm eff}$. Moreover, the tip of the CS lobe is given by $J_{\rm c}=(V_1^{\rm eff}-V_2^{\rm eff})/2$. It is worth noting that the MF energy of Eq.~(\ref{EneChDip}) is valid not only for the $V_{\rm dip}$ model, but generally for systems with checkerboard sublattice structure regardless of the form of $V_{jl}$.

\begin{figure}[t]
\includegraphics[scale=0.8]{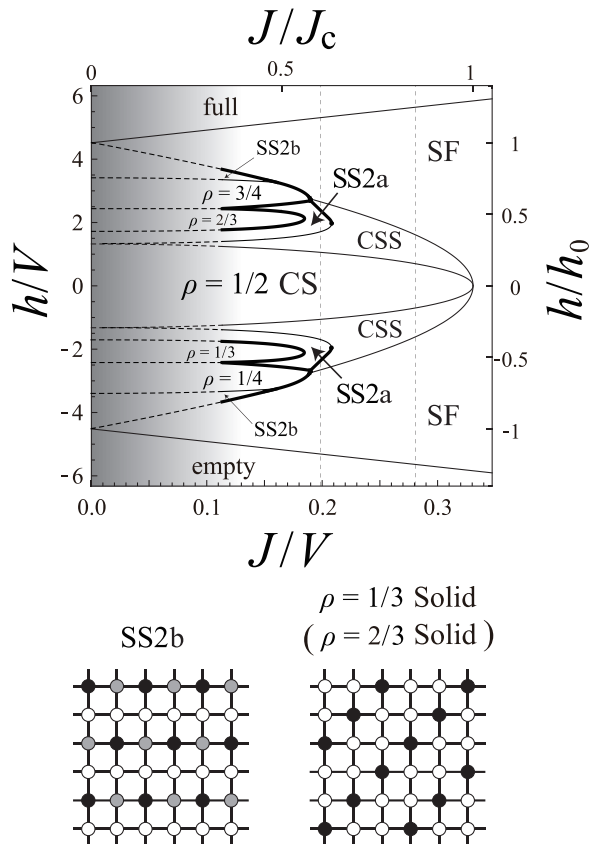}
\caption{\label{MFPDDip}
The same as in Fig.~\ref{MFPDV1V2} for the $V_{\rm dip}$ model. The $\rho=1/4$ ($3/4$) solid state here has the b-type symmetry in Fig.~\ref{sublatticesV1V2}(II). The lower panels are the sketches of the SS2b and $\rho=1/3$ ($\rho=2/3$) solid states. Many other phases with more complex structure can emerge in the small $J/V$ (shaded) region. The dashed vertical lines mark the location of $J/J_{\rm c}=0.6$ and $0.85$.}
\end{figure}

The resulting MF phase diagram shown in Fig.~\ref{MFPDDip} has a similar structure to that of the $V_1$-$V_2$ model in Fig.~\ref{MFPDV1V2}, especially for the region of $J/V>0.2$. 
However, many additional phases emerge for the region of smaller $J/V$ due to the long-ranged character of the dipole-dipole interaction. 
Within our analysis (see Appendix~\ref{appA}), we found the supersolid, named SS2b, and the solid phases with $\rho=1/3$ and $2/3$ in addition to the phases appearing in the $V_1$-$V_2$ model. Unlike the $V_1$-$V_2$ model, the two possible structures of the $\rho=1/4$ ($3/4$) solid state shown in Fig.\ref{sublatticesV1V2}(II) can be distinguished even within the MF theory; the b-type structure has lower energy. The emergence of these solid phases is consistent with the QMC results.~\cite{sansone-10} 
Although many other phases can emerge for smaller $J/V$, we do not extend the calculations to more complex sublattice structures, since our main focus is the stability of the CSS phase.

It should be noted that the ratio of the effective NNN interaction strength to the NN one is fixed in the $V_{\rm dip}$ model as$V_2^{\rm eff}/V_1^{\rm eff}\approx0.55$. This value obviously exceeds the limit $V_2/V_1=1/2$, above which the striped phases emerge in place of the checkerboard ones in the case of the $V_1$-$V_2$ model. Nevertheless, we have to keep in mind that the effective interactions $V_{1,2}^{\rm eff}$ are made by the summation of the long-range interactions between various pairs with different distances. Therefore, the limit predicted for the $V_1$-$V_2$ model cannot be directly applied to the $V_{\rm dip}$ model.

In the $V_{\rm dip}$ model, the MF energy per site for the striped phases is written as
\begin{eqnarray}
E_{0}^{(\rm st)}/M&=&-JS^2 \left(\sin \theta_{\rm R_1}+\sin \theta_{\rm R_2}\right)^2\nonumber\\
&&+\tilde{V}_1^{\rm eff}S^2 \left(\cos \theta_{\rm R_1}+\cos \theta_{\rm R_2}\right)^2/2\nonumber\\
&&+2\tilde{V}_2^{\rm eff}S^2\cos \theta_{\rm R_1}\cos \theta_{\rm R_2}\nonumber\\
&&-hS(\cos \theta_{\rm R_1}+\cos \theta_{\rm R_2})/2.
\label{EneStDip}
\end{eqnarray}
This expression is formally equivalent to that of the $V_1$-$V_2$ model [Eq.~(\ref{EneSt})] with the effective interactions 
\begin{subequations}
\begin{eqnarray}
\tilde{V}_1^{\rm eff}&\equiv& \frac{2}{z}\sum_{l_{\rm R_1}} V_{j_{\rm R_1}l_{\rm R_1}}= 2.025V,\\
\tilde{V}_2^{\rm eff}&\equiv&\frac{1}{z}\sum_{l_{\rm R_1}} \left( V_{j_{\rm R_2}l_{\rm R_1}}- V_{j_{\rm R_1}l_{\rm R_1}}\right)= 0.2339V.
\end{eqnarray}\label{VeffST}\end{subequations}
However, the effective NN and NNN interactions have different values for the checkerboard ($V_{1,2}^{\rm eff}$) and striped ($\tilde{V}_{1,2}^{\rm eff}$) phases [compare Eqs.~(\ref{VeffCH}) and~(\ref{VeffST})]. Hence, the large value of $V_2^{\rm eff}/V_1^{\rm eff}$ in the checkerboard phases does not mean that the striped phases are energetically preferred, and the checkerboard order is always favored over the striped one in the $V_{\rm dip}$ model. As an example, we show the comparison of the MF energies of the CS and striped solid states: 
\begin{subequations}
\begin{eqnarray}
E_{0}^{(\rm ch)}/M&=&-2S^2(V_1^{\rm eff}-V_2^{\rm eff})\nonumber\\
&=&-0.3307V~~ ({\rm CS}) ,\\
E_{0}^{(\rm st)}/M&=&-2S^2\tilde{V}_2^{\rm eff}\nonumber\\
&=&-0.1169V~~ ({\rm striped}~{\rm solid}) . 
\end{eqnarray}\end{subequations}

According to Eqs.~(\ref{hcV1V2}), the CSS region, $h_{{\rm c}1}<|h|<h_{{\rm c}2}$, gets wider for a larger value of $V_2/V_1$ ($V_2^{\rm eff}/V_1^{\rm eff}$). In the $V_{\rm dip}$ model, the ratio $V_2^{\rm eff}/V_1^{\rm eff}\approx 0.55$ is larger than $V_2/V_1$ of the $V_1$-$V_2$ model with the checkerboard order, the CSS region is also larger in the MF level. This is one of the two main reasons why the CSS phase is stable in the case of the dipole-dipole interactions. Comparing the width in units of $h_0$, for example, at $J/J_{\rm c}=0.85$ in Figs.~\ref{MFPDV1V2}(a), \ref{MFPDV1V2}(b), and~\ref{MFPDDip}, we indeed see that the $V_{\rm dip}$ model has a wider region of the CSS phase than the $V_1$-$V_2$ model. Moreover, despite the large value of $V_2^{\rm eff}/V_1^{\rm eff}$, the SS2a region in Fig.~\ref{MFPDDip} is relatively suppressed compared with that in Fig.~\ref{MFPDV1V2}(b). This means that the direct competition of the checkerboard and striped density-wave orders is much weaker than the case of the $V_1$-$V_2$ model. The suppression of the competition can be also seen in the excitation spectra, which will be discussed in the next section.

\section{\label{4}Linear spin-wave analysis}
In this section, we discuss the strength of quantum fluctuations around the MF ground states within the linear spin-wave (LSW) theory.~\cite{pich-98,scalettar-95,coletta-12} First, we perform local rotations of the spin reference frame in Eq.~(\ref{spinH}), so that the new spin quantization axis is oriented along the direction of the classical pseudospin vector: 
\begin{eqnarray}
\left(
\begin{array}{c}
\hat{S}_j^x \\
\hat{S}_j^y \\
\hat{S}_j^z 
\end{array}
\right)&=&\left(
\begin{array}{ccc}
\cos \theta_j & 0 & \sin \theta_j \\
0 & 1 & 0 \\
-\sin \theta_j & 0 & \cos \theta_j
\end{array}
\right) \left(
\begin{array}{c}
\tilde{S}_j^x \\
\tilde{S}_j^y \\
\tilde{S}_j^z 
\end{array}
\right). \label{rotation}
\end{eqnarray}
Furthermore, we introduce new bosonic variables via the Holstein-Primakoff transformation, 
\begin{subequations}
\begin{eqnarray}
\tilde{S}_j^z&=&S-\hat{b}_j^\dagger\hat{b}_j, \\
\tilde{S}_j^x&=&\frac{1}{2}\left(\sqrt{\mathstrut 2S-\hat{b}_j^\dagger\hat{b}_j}\hat{b}_j+\hat{b}_j^\dagger\sqrt{\mathstrut 2S-\hat{b}_j^\dagger\hat{b}_j}\right),\\
\tilde{S}_j^y&=&\frac{1}{2i}\left(\sqrt{\mathstrut 2S-\hat{b}_j^\dagger\hat{b}_j}\hat{b}_j-\hat{b}_j^\dagger\sqrt{\mathstrut 2S-\hat{b}_j^\dagger\hat{b}_j}\right),
\end{eqnarray}\label{HPtr}\end{subequations}
to describe quantum fluctuations around the classical spin angles. Within the LSW approximation, we keep the terms up to the second order in the boson operators:
\begin{eqnarray}
\hat{H}_{\rm spin}\approx E_{0}+\hat{H}_{2},
\end{eqnarray}
where $E_{0}$ is identical to the MF energy given by Eq.~(\ref{eq:EneCl}). The linear term in boson operators disappears by substituting the MF solutions into $\theta_j$. Diagonalizing $\hat{H}_{2}$, we calculate the LSW excitation spectra $\omega ({\bf q})$ and the number of ``spin waves'' $\langle \hat{b}_j^\dagger \hat{b}_j\rangle$ to estimate the strength of the quantum fluctuations (see Appendix~\ref{appB} for details). By calculating the number of spin waves, one can roughly estimate the strength of quantum fluctuations around the MF solutions obtained in Sec.~\ref{3}. In the spin language, the value of $\langle \hat{b}_j^\dagger \hat{b}_j\rangle$ corresponds to the spin reduction from its classical value $S$ due to the zero-point fluctuations.

We will show the results of the excitation spectra in Sec.~\ref{4-1} and of the the number of spin waves in Sec.~\ref{4-2}.
For the $V_1$-$V_2$ model, the LSW excitation spectra of SF, CS, and CSS states have already been discussed in detail in Ref.~\onlinecite{scalettar-95}, and it was confirmed that the softening of roton excitations causes the phase transition from the SF to CSS state. As for the $V_{\rm dip}$ model, although we used in Ref.~\onlinecite{danshita-10} the LSW theory to discuss the critical velocity of flowing CSS states, detailed results of the spectra have not been presented yet. Moreover, to date, no studies estimating the strength of quantum fluctuations from the values of $\langle \hat{b}_j^\dagger \hat{b}_j\rangle$ have been demonstrated for the comparison of the two models.

\subsection{\label{4-1}The excitation spectra}
As mentioned above, the LSW excitation spectra for the $V_1$-$V_2$ model have already been analyzed in Ref.~\onlinecite{scalettar-95}. Hence, we show here only the case of the dipole-dipole interaction given in Eq.~(\ref{Dip}). In the calculations, we have to take an infinite summation in Eq.~(\ref{Vq}) due to the long-range nature. To avoid the practical difficulty, we introduce a cutoff distance on the dipole-dipole interaction as $V_{jl}=0$ for $\left|{\bf r}_{j}-{\bf r}_{l}\right|>16d$ only in this section (namely, in Secs.~\ref{4-1} and~\ref{4-2}). For this truncated dipole-dipole interaction, the values of the effective NN and NNN interactions are $V_1^{\rm eff}=1.410V$ and $V_2^{\rm eff}=0.7495V$ and the ratio is $V_2^{\rm eff}/V_1^{\rm eff}\approx 0.53$. 

\begin{figure}[t]
\includegraphics[scale=0.5]{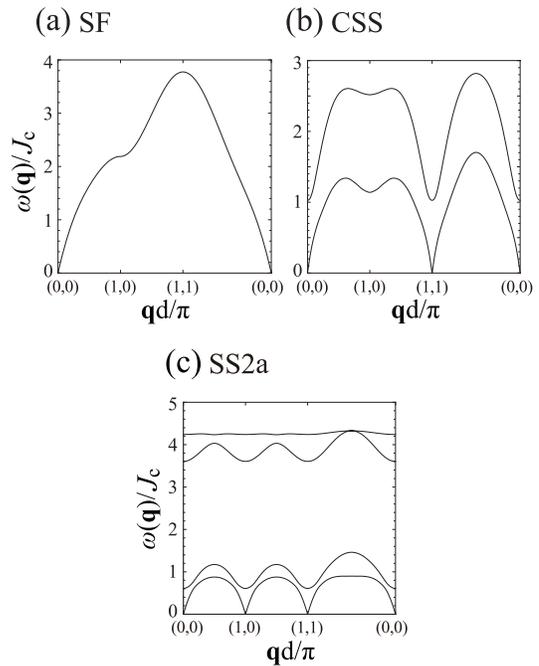}
\caption{\label{SW}
Excitation spectra $\omega ({\bf q})$ of the $V_{\rm dip}$ model in the (a) SF (at $h/h_0=\pm 1$), (b) CSS (at $h/h_0=\pm 0.58$), and (c) SS2a (at $h/h_0=\pm 0.5$) phases for $J/J_{\rm c}=0.6$ .
}
\end{figure}

Solving Eq.~(\ref{GF}), we plot in Figs.~\ref{SW}(a)-\ref{SW}(c) the excitation spectra $\omega ({\bf q})$ for the SF, CSS, and SS2a phases along the line $J/J_{\rm c}=0.6$ marked in Fig.~\ref{MFPDDip}. The excitation spectra have a Nambu-Goldstone mode reflecting the spontaneous breaking of the U(1) symmetry. The spectrum of the CSS phase consists of two branches due to the two-sublattice structure, and the lower branch has gapless, linear dispersions around ${\bf q}=(0,0)$ and $(\pi /d,\pi /d)$, which is the ordering vector of the checkerboard phases. In the SS2a phase, in which the checkerboard and stripe orders coexist, the lowest mode is gapless at ${\bf q}=(\pi /d,0)$ in addition to at ${\bf q}=(0,0)$ and $(\pi /d,\pi /d)$. 
\begin{figure}[t]
\includegraphics[scale=0.5]{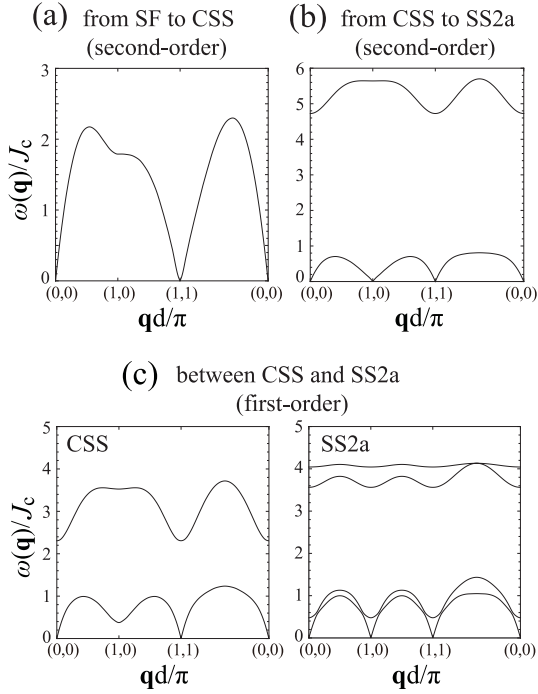}
\caption{\label{SWb}
Excitation spectra $\omega ({\bf q})$ of the $V_{\rm dip}$ model at the second-order transition (a) from the SF to CSS phase (at $h=\pm h_{{\rm c}2}$) and (b) from the CSS to SS2a phase (at $h/h_0=\pm 0.3819$) for $J/J_{\rm c}=0.6$. (c) The same as in panels (a) and (b) at the first-order transition between the CSS and SS2a phases (at $h/h_0=\pm 0.5348$). The CSS and SS2a states are energetically degenerate at this point. }
\end{figure}

Figures~\ref{SWb}(a)-\ref{SWb}(d) show the excitation spectra at the phase transitions between the different phases. When one approaches the CSS phase from the SF region, a roton-like minimum at ${\bf q}=(\pi /d,\pi /d)$ develops, and it touches zero at the boundary $h=\pm h_{{\rm c}2}$ as shown in Fig.~\ref{SWb}(a), causing the second-order phase transition to the CSS state. In a similar way, a roton-like mode at ${\bf q}=(\pi /d,0)$ causes the second-order transition from CSS to SS2a [see, Fig.~\ref{SWb}(b)]. In contrast, as shown in Fig.~\ref{SWb}(c), such a signal does not appear in the spectra at the first-order phase transitions. The above-mentioned properties of the excitations qualitatively agree with the case of the $V_1$-$V_2$ model.

\subsection{\label{4-2}The number of spin waves}
For all the cases of Figs.~\ref{MFPDV1V2}(a), \ref{MFPDV1V2}(b), and \ref{MFPDDip}, the system exhibits the phase transition from SF to CSS, and then it reaches the CS phase if the value of $h/h_0$ increases from a negative value to zero along the line of $J/J_{\rm c}=0.85$. We will plot the number of spin waves $\langle \hat{b}_j^\dagger \hat{b}_j\rangle$ along this line as a function of the filling factor $\rho$. Within the MF analysis, the filling factor $\rho$ shows a linear increase with the chemical potential $h$ both in the SF and CSS phases, and the slope of the line, which is proportional to the compressibility, is always larger in the CSS phase than in the SF phase:
\begin{eqnarray}
\rho=\left\{ \begin{array}{ll}
1/2+ h/2h_{{\rm c}3}& ~({\rm SF})  \\
1/2+ \left(h\mp h_{{\rm c}1}\right)/16V_2S& ~({\rm CSS})

\end{array} \right.,\label{FF}
\end{eqnarray}
where the upper (lower) signs are for positive (negative) values of $h$. Recall that $V_2$ has to be replaced with $V_2^{\rm eff}$ for the $V_{\rm dip}$ model. It should be noted that the critical filling factor $\rho_{\rm c}$ at the SF-CSS transition point, which is obtained by substituting $h=\pm h_{{\rm c}2}$ into Eq.~(\ref{FF}), can be expressed as a function only of $J/J_{\rm c}$: 
\begin{eqnarray}
\rho_{\rm c}=\frac{1}{2}\pm\frac{1}{2}\sqrt{\frac{1-J/J_{\rm c}}{1+J/J_{\rm c}}}. 
\end{eqnarray}
It takes $\rho_{\rm c}=0.3576$ for $J/J_{\rm c}=0.85$ in the low-density (negative $h$) side.

\begin{figure}[b]
\includegraphics[scale=0.5]{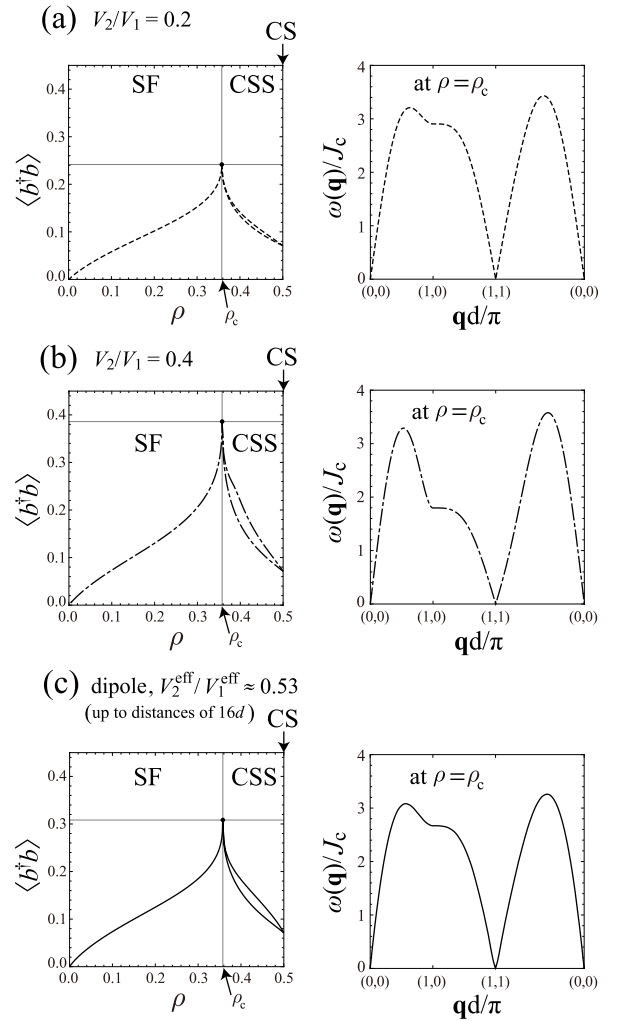}
\caption{\label{NUMofSW}
The number of spin waves $\langle \hat{b}_{j_\alpha}^\dagger \hat{b}_{j_\alpha}\rangle$ as a function of the filling factor $\rho$ for $J/J_{\rm c}=0.85$ and the excitation spectra $\omega_{\bf q}$ at the SF-CSS phase transition point $\rho=\rho_{\rm c}$. The panels (a), (b), and (c) show the results for the $V_1$-$V_2$ model with $V_2/V_1=0.2$ and $0.4$, and for the $V_{\rm dip}$ model with the truncated dipole-dipole interaction, respectively. In the CSS phase, $\langle \hat{b}_{j_\alpha}^\dagger \hat{b}_{j_\alpha}\rangle$ takes two different values on each sublattice. }
\end{figure}
Figures~\ref{NUMofSW}(a)-\ref{NUMofSW}(c) show the results for the $V_1$-$V_2$ model with $V_2/V_1=0.2$ and $0.4$, and for the $V_{\rm dip}$ model. In all the cases, we can see that $\langle \hat{b}_{j}^\dagger \hat{b}_{j}\rangle$ has a peak at the SF-CSS phase transition, which means that quantum fluctuations are particularly strong at the phase boundary. The number of spin waves for $V_2/V_1=0.4$ in Fig.~\ref{NUMofSW}(b) is much larger than the case of $V_2/V_1=0.2$ in Fig.~\ref{NUMofSW}(a). This is attributed to the strong competition of the NN and NNN interactions. In fact, as shown in the right panel of Fig.~\ref{NUMofSW}(b), the excitation spectrum $\omega_{\bf q}$ exhibits a remarkable drop at ${\bf q}=(\pi /d,0)$, which indicates the existence of strong striped density-wave fluctuations. The maximum value of $\langle \hat{b}_{j}^\dagger \hat{b}_{j}\rangle$ for $V_2/V_1=0.4$ reaches about 77 percent of the classical value of the spin length $S=1/2$, which means that the predictions for the ground states within the MF theory are unreliable. Actually we will show in Sec.~\ref{5} that the CSS phase predicted by the MF theory almost completely disappears due to the strong quantum fluctuations.

On the other hand, the excitation spectrum $\omega_{\bf q}$ for the $V_{\rm dip}$ model in the right panel of Fig.~\ref{NUMofSW}(c) does not exhibit a significant drop at ${\bf q}=(\pi /d,0)$ unlike the case of $V_2/V_1=0.4$. This comes from the fact that $V_2^{\rm eff}$ is not just the NNN interaction but the summation of various long-range interactions that weakens the competition with the stripe order. Therefore, as shown in Fig.~\ref{NUMofSW}(c), the quantum fluctuations in the $V_{\rm dip}$ model are relatively weak for its large value of $V_2^{\rm eff}/V_1^{\rm eff}$. This is the second reason for the stability of the CSS state in the $V_{\rm dip}$ model. The CSS region predicted by the MF theory is significantly reduced by the quantum fluctuations but still remains sufficiently large (see Ref.~\onlinecite{sansone-10} and Sec.~\ref{5} of this paper).

\section{\label{5}Large-size cluster mean-field method and scaling analysis}
We drew the ground-state phase diagram of the two models within the MF theory in Figs.~\ref{MFPDV1V2}(a),~\ref{MFPDV1V2}(b), and~\ref{MFPDDip}. However, according to the previous section, the fluctuations around the classical ground states are too large to completely ignore in any case. In this section, considering the MF results obtained in Sec.~\ref{3} as a starting point, we discuss how the quantum fluctuations change the features of the ground-state phase diagrams by employing a large-size CMF method.~\cite{yamamoto-12} We perform the calculations based on rectangular-shaped clusters, and then extrapolate the results with respect to the cluster size. The obtained results will be compared with the QMC data in Ref.~\onlinecite{sansone-10} for the $V_{\rm dip}$ model. As for the $V_1$-$V_2$ model, although the authors of Ref.~\onlinecite{batrouni-00} concluded that the CSS state is thermodynamically unstable from the QMC calculations for $V_1=3J$, the entire phase diagram including the effects of quantum fluctuation has not been produced yet. We will confirm, in the $(J/V_1,h/V_1)$-plane, that the CSS phase is almost completely destroyed by the strong quantum fluctuations in the $V_1$-$V_2$ model. Hereafter, we will use the full (untruncated) dipole-dipole interaction again in the $V_{\rm dip}$ model.

\subsection{\label{5-1}The CMF method}
First, we describe the details of our CMF approach.~\cite{yamamoto-12} The standard MF theory approximates the system by single-site problems in effective fields. A natural extension of the single-site approximation is the use of ``clusters'' of multiple sites as an approximate system.~\cite{bethe-35,peierls-36,weiss-48,campbell-72,du-03,etxebarria-04,neto-06,oguchi-55,buonsante-04,mcintosh-12} For example, the Bethe-Peierls-Weiss (BPW) method~\cite{bethe-35,peierls-36,weiss-48} employs a cluster consisting of one central site and its directly connected sites, e.g., a cluster of $(1+6+6)$ sites for a triangular-lattice system with NN and NNN interactions.~\cite{campbell-72} Treating exactly the interactions within the cluster, one can partially take into account the effects of correlations between particles (or spins). However, the BPW method and its extensions~\cite{du-03,etxebarria-04} cannot be applied to an infinite-range interaction model like the $V_{\rm dip}$ model since all sites are ``directly connected'' by the long-range interactions.

Oguchi's method~\cite{oguchi-55} is another simple way to extend the MF theory to clusters. In Ref.~\onlinecite{oguchi-55}, Oguchi studied ferromagnetism and antiferromagnetism of the low-dimensional Heisenberg model by using a cluster of up to three spins to include the short-range correlations between the spins. Since the influence from the spins outside of the cluster is also included as effective internal fields, we can treat even a system with infinite-range interactions. However, as Oguchi himself pointed out, the cluster of two or three sites is too small to sufficiently take into account the effects of the correlations (or quantum fluctuations).

\begin{figure}[t]
\includegraphics[scale=0.6]{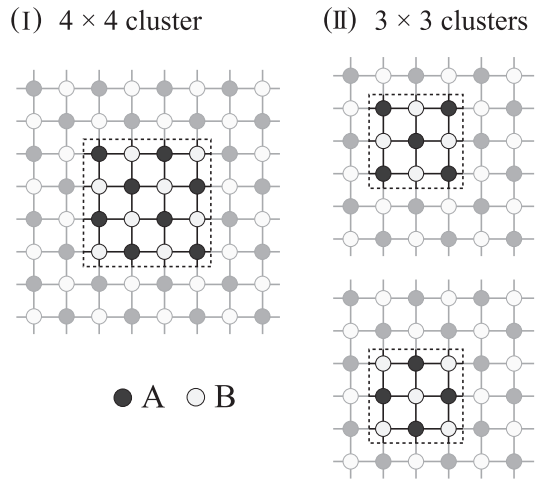}
\caption{\label{ExEmbed}
Clusters of (I) $4\times 4$ sites and (II) $3\times 3$ sites embedded into the checkerboard sublattice pattern. }
\end{figure}

Our CMF approach~\cite{yamamoto-12} is an extension of Oguchi's method to larger-size clusters and to multiple-sublattice problems. Although we use here the pseudospin form of the Hamiltonian, Eq.~(\ref{spinH}), to explain the procedure of our method, the same manner can be applied straightforwardly to hardcore bosons and even to softcore-boson models. First, we assume a sublattice structure expected to emerge in the parameter range. Then, we embed a cluster of $N_{\rm C}$ sites into the background sublattice structure. Figures~\ref{ExEmbed}(I) and \ref{ExEmbed}(II) show, as examples, the cases of $N_{\rm C}=4\times 4$ and $N_{\rm C}=3\times 3$, which we refer to as CMF-$4\times 4$ and CMF-$3\times 3$ under the assumption of the checkerboard sublattice structure. 
As for the case of the CMF-$3\times 3$, we have two inequivalent choices for embedding the cluster. We have to deal with both of them equally as in the BPW method for multiple-sublattice systems.~\cite{neto-06} Now, instead of treating the many-body problem in the whole system given by Eq.~(\ref{spinH}), we consider the effective cluster Hamiltonian $H_C$ written as the following general form: 
\begin{eqnarray}
\hat{H}_C&=&
-2J\sum_{\langle j,l \rangle\in C}
\left(\hat{S}^{x}_{j} \hat{S}^{x}_{l}+\hat{S}^{y}_{j} \hat{S}^{y}_{l}\right)+\frac{1}{2}\sum_{ j,l\in C}V_{jl} \hat{S}^{z}_{j}\hat{S}^{z}_{l}\nonumber\\
&&
- \sum_{j\in C}\left(h+h_j^{z,{\rm eff}} \right) \hat{S}^{z}_j- \sum_{j\in C}h_j^{x,{\rm eff}} \hat{S}^{x}_j,
\label{spinH_C}
\end{eqnarray}
in which the interactions within the cluster are treated exactly, while the interactions between the spins in the cluster and the rest of the system are approximately included via the effective fields 
\begin{subequations}\begin{eqnarray}
h_j^{z,{\rm eff}}&\equiv& -\sum_{l\in\bar{C}}V_{jl} m^{z}_{l},\\
h_j^{x,{\rm eff}}&\equiv& 2\sum_{l\in\bar{C}}J_{jl} m^{x}_{l},
\end{eqnarray}\end{subequations}
where $\bar{C}$ is the part of the system outside the cluster and $m^{z,x}_{l}\equiv\langle \hat{S}_l^{z,x}\rangle_{\rm CMF}$ are the expectation values within the CMF method, which act as the mean fields from the spins in $\bar{C}$. Here, we chose again the $xz$ plane as the plane in which the spins lie; i.e., $\langle \hat{S}^{y}_j\rangle=0$. If we have two or more possibilities of choosing the cluster like in the CMF-$3\times 3$, we should consider all the corresponding cluster Hamiltonians like $\hat{H}_{C_1}$, $\hat{H}_{C_2}$, $\cdots$.

Note that in our CMF method, we consider the $N_{\rm C}$-site problem in the cluster just as a reference system to estimate the values of the mean fields $m^{z,x}_{l}$, which depend only on the background sublattice index of the site; i.e., $m^{z,x}_{\alpha}\equiv m^{z,x}_{l_\alpha}$. For example, the effective fields acting on the top-left site ``$1$'' in the $4\times 4$ cluster of Fig.~\ref{ExEmbed}(I) can be written as the following explicit forms:
\begin{eqnarray}\begin{array}{rcl}
h_1^{z,{\rm eff}}&=& -2V_1 m^{z}_{\rm B}-3V_2 m^{z}_{\rm A},\\
h_1^{x,{\rm eff}}&=& 4J m^{x}_{\rm B}
\end{array}\label{hzx1V1V2}\end{eqnarray}
for the $V_1$-$V_2$ model and 
\begin{eqnarray}\begin{array}{rcl}
h_1^{z,{\rm eff}}&=&\displaystyle -\left(4V_1^{\rm eff}-2V-\frac{2V}{\sqrt{5}^3}-\frac{2V}{3^3}-\frac{2V}{\sqrt{13}^3}\right) m^{z}_{\rm B}\\
&&\displaystyle -\left(4V_2^{\rm eff}-\frac{V}{\sqrt{2}^3}-\frac{2V}{2^3}-\frac{V}{\sqrt{8}^3}-\frac{2V}{\sqrt{10}^3}\right.\\
&&\displaystyle \left.-\frac{V}{\sqrt{18}^3}\right) m^{z}_{\rm A},\\
h_1^{x,{\rm eff}}&=& 4J m^{x}_{\rm B}
\end{array}\label{hzx1Vdip}\end{eqnarray}
for the $V_{\rm dip}$ model (see Appendix~\ref{appC} for more details). 
The values of the mean fields $m^{z}_{\alpha}$ and $m^{x}_{\alpha}$ are calculated self-consistently as the expectation values of the pseudospins inside the cluster as follows: 
\begin{eqnarray}
m^{z,x}_{\alpha}&=&\langle \hat{S}_{j_\alpha}^{z,x}\rangle_{\rm CMF}\nonumber\\
&=&\frac{1}{M_{\rm C}N_{\alpha}}\sum_{n}\sum_{j_\alpha\in C_n} \frac{{\rm Tr}\left(\hat{S}_{j_\alpha}^{z,x}  e^{-\beta H_{C_n}}\right)}{{\rm Tr}\left(e^{-\beta H_{C_n}}\right)} \label{SCEq}
\end{eqnarray}
where $\beta=1/T$ (we take $T\rightarrow 0$ in this paper), $M_{\rm C}$ is the number of the possible choices of the cluster, and $N_{\alpha}\equiv \sum_n N_{\alpha,n}$ is the summation of the number of sites belonging to sublattice $\alpha$ in cluster $C_n$. For example, $M_{\rm C}=2$ in the CMF-$3\times 3$ for checkerboard phases in Fig.~\ref{ExEmbed}(II) and we have $(N_{{\rm A},1},N_{{\rm B},1})=(5,4)$ and $(N_{{\rm A},2},N_{{\rm B},2})=(4,5)$ for the two clusters $C_1$ and $C_2$, respectively, which leads to $N_{\rm A}=N_{\rm B}=9$.

This method reduces to the conventional MF (namely, Weiss's molecular-field) theory for $N_{\rm C}=1$, and becomes exact in the limit $N_{\rm C}\rightarrow \infty$. We have to diagonalize the cluster Hamiltonian to take the trace on the right-hand side of the self-consistent equation, Eq.~(\ref{SCEq}). Thus the practical limit of the cluster size $N_{\rm C}$ is determined by the largest number of sites which can be treated by exact diagonalization techniques. It should be noted, however, that some of the symmetries of the original Hamiltonian, Eq.~(\ref{spinH}), are broken in the effective cluster Hamiltonian due to the existence of mean fields.

\subsection{\label{5-2}The CMF results for the $4\times 4$ cluster}
\begin{figure}[t]
\includegraphics[scale=0.5]{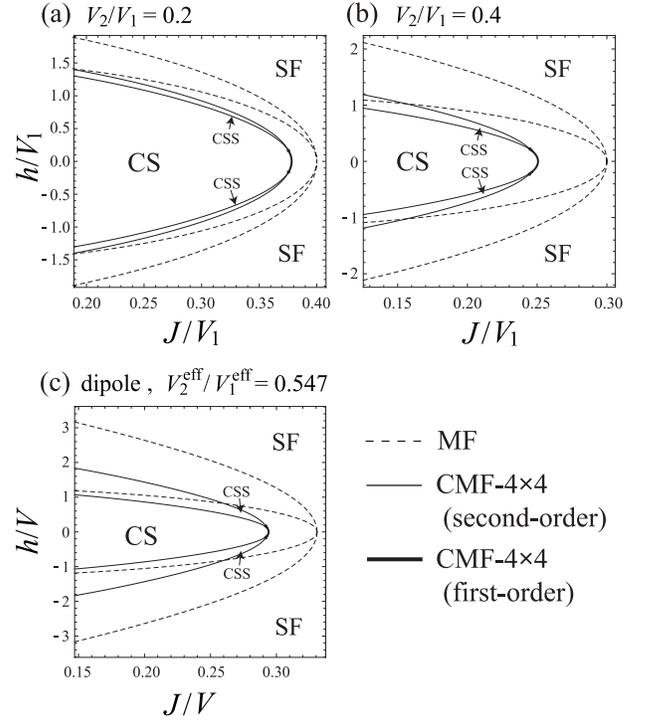}
\caption{\label{CMF16V}
Phase boundaries between the CS, CSS, and SF phases by the CMF-$4\times 4$ calculations for the $V_1$-$V_2$ model with (a) $V_2/V_1=0.2$ and (b) $V_2/V_1=0.4$, and for (c) the $V_{\rm dip}$ model. The thin and thick solid lines indicate the second-order and first-order transitions. For comparison, the MF results, Eq.~(\ref{hcV1V2}), are shown by the dashed lines. }
\end{figure}
First, we show the results of the CMF-$4\times4$ method for the two models and compare them with the MF results. We focus on the checkerboard phases and discuss the influence of quantum fluctuations on the locations of the phase boundaries between the CS, CSS, and SF phases, namely $h=\pm h_{{\rm c}1}$ and $\pm h_{{\rm c}2}$ in Eqs.~(\ref{hcV1V2}). Note that there are no quantum fluctuations at the boundary between the SF and MI phases, $h=\pm h_{{\rm c}3}$, and thus the expression in Eq.~(\ref{hcV1V2}c) does not change for any $N_{\rm C}$. Other phases with more complex symmetries, such as three-sublattice and four-sublattice phases, are affected more strongly by the quantum fluctuations, and the locations are shifted towards the region of much smaller values of $J/J_{\rm c}$ than those of the MF theory.~\cite{sansone-10,ng-10} Thus, we leave them out of the scope of the rest of this paper, treating only a relatively large-$J/J_{\rm c}$ region.

For the checkerboard phases, we can obtain the SF order parameter $|\Psi|=|m^{x}_{\rm A}+m^{x}_{\rm B}|/2$, the CS order parameter $|\rho_{\bf Q}|=|m^{z}_{\rm A}-m^{z}_{\rm B}|/2$, and the filling factor $\rho=1/2+(m^{z}_{\rm A}+m^{z}_{\rm B})/2$ by solving Eq.~(\ref{SCEq}) self-consistently. The second-order transition boundary from CSS to CS is determined by the point at which $|\Psi|$ vanishes (or at which the value of $\rho$ reaches $1/2$). On the other hand, the CS order parameter $|\rho_{\bf Q}|$ vanishes (namely, $m^{z}_{\rm A}=m^{z}_{\rm B}$) at the second-order transition from CSS to SF. In addition to these second-order (continuous) transitions, we find that first-order (discontinuous) transitions between CS and SF and between CSS and SF can also appear due to the effects of the quantum fluctuations. In the CMF formalism, we cannot directly calculate the value of free energy of the system. Instead, we use the Maxwell construction in the ($J,\chi$) plane to determine the first-order phase boundaries. The quantity $\chi$ is defined by $\chi\equiv \sum_{\langle j,l \rangle}\langle\hat{a}^{\dagger}_{j} \hat{a}_{l}+\hat{a}^{\dagger}_{l} \hat{a}_{j}\rangle/M$.

\begin{figure}[t]
\includegraphics[scale=0.46]{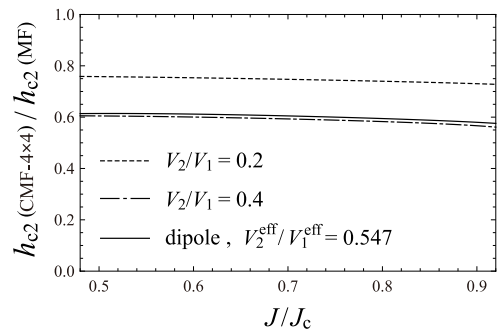}
\caption{\label{MFvs16}
The ratio of $h_{{\rm c}2}$ obtained by the CMF-$4\times 4$ method to the MF value at fixed $J/J_{\rm c}$. }
\end{figure}

Figures~\ref{CMF16V}(a)-\ref{CMF16V}(c) show the phase boundaries between the CS, CSS, and SF phases obtained by the CMF-$4\times 4$ method with the corresponding MF results. In all the cases, the regions of the CS and CSS phases shrink considerably because of the quantum fluctuations. The reductions of the values of $J_{\rm c}$ from the MF values are $5.5$ percent for $V_2/V_1=0.2$, $16.5$ percent for $V_2/V_1=0.4$, and $11.1$ percent for the $V_{\rm dip}$ model within the CMF-$4\times 4$ level. This fact indicates that the quantum fluctuation of the $V_1$-$V_2$ model is stronger for a larger value of $V_2/V_1$, and that of the $V_{\rm dip}$ model is small relative to the large value of $V_2^{\rm eff}/V_1^{\rm eff}$, which is consistent with the spin-wave prediction in Sec.~\ref{4}. In Fig.~\ref{MFvs16}, we compare the reductions of the value of $|h|$ at the CSS-SF boundary, $h_{{\rm c}2}$, from the MF value given in Eq.~(\ref{hcV1V2}b). 
This result also confirms the statement made in Sec.~\ref{4}.

\begin{figure}[t]
\includegraphics[scale=0.46]{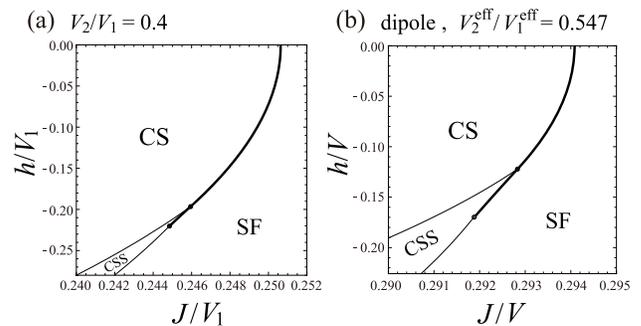}
\caption{\label{EnlargedView}
Enlarged views of the region around the tip of the lobe in Fig.~\ref{CMF16V}(b) [panel (a)] and Fig.~\ref{CMF16V}(c) [panel (b)]. The first-order transitions from the CS to SF phase and from the CSS to SF phase are found. Also in the case of $V_2/V_1=0.2$, we find a narrow but finite region where the first-order transitions occur. }
\end{figure}

Other than the shrinking of the CS and SS phases, we can see a qualitative difference between the CMF and MF results. As clearly seen in the enlarged views, Figs.~\ref{EnlargedView}(a) and~\ref{EnlargedView}(b), the direct first-order (discontinuous) transition from the CS to the SF phase emerges in the CMF result, and the transition between the CSS and SF phase also becomes discontinuous near the triple point of the CS, CSS, and SF phases. This result is attributed to the fact that the classical degeneracy of the CS, CSS, and SF states at $J=J_{\rm c}$ and $h=0$ is lifted by taking into account the quantum fluctuations. However,  the first-order transitions occur in a narrow range of parameters and the discontinuity of the order parameter is very small, and thus the first-order nature has not been reported previously in the QMC works.~\cite{batrouni-00,sansone-10} 

\subsection{\label{5-3}The cluster-size scaling}
We perform the infinite-size extrapolation, $N_{\rm C}\rightarrow \infty$, of the CMF results with different-size clusters. 
\begin{table}[b]
\caption{\label{table1}A series of clusters used in our CMF calculations. The values of $N_{\rm B}$ and $\lambda$ are also listed. }
\includegraphics[scale=0.5]{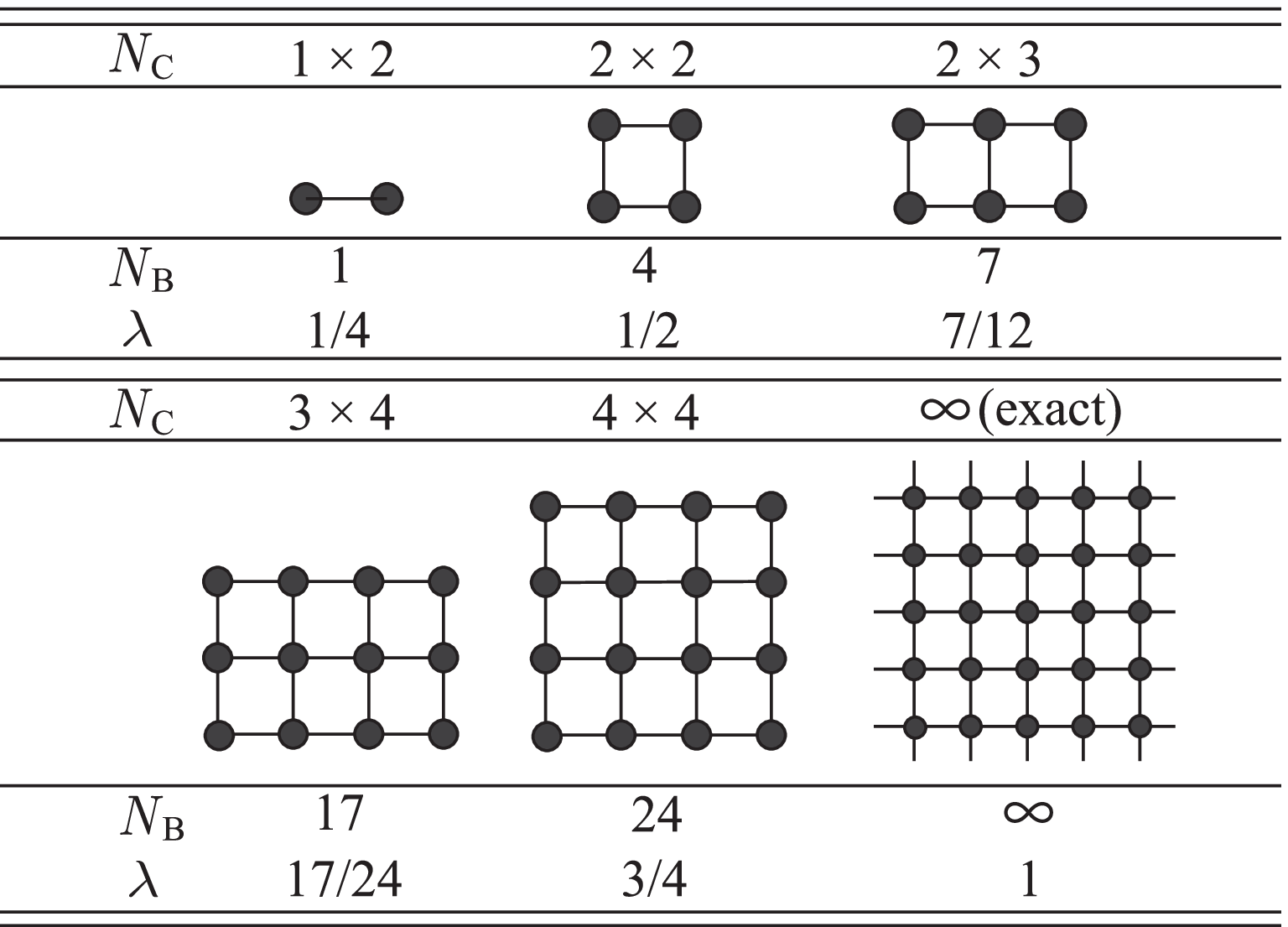}
\end{table}
We use a series of rectangular-shaped clusters of $N_{\rm C}=1\times 2$, $2\times 2$, $2\times 3$, $3\times 4$, and $4\times 4$, which are shown in Table~\ref{table1}. The clusters with odd numbers of sites, e.g., $N_{\rm C}=3\times 3$, are not treated here, because they may belong to a different scaling series from that of the clusters with even numbers of sites. To perform the infinite-size extrapolation, we introduce the scaling parameter $\lambda$ defined by $\frac{N_{\rm B}}{N_{\rm C}\times z/2}$, which varies from $0$ to $1$. Here, $N_{\rm B}$ is the number of bonds within the cluster and the denominator means the number of bonds of the original lattice per $N_{\rm C}$ sites. The parameter $\lambda$ provides an indication of how much the correlation effects between the particles are taken into account by using the cluster. The value of $\lambda$ for each cluster is listed in Table~\ref{table1}. Note that the MF ($N_{\rm c}=1$) and exact ($N_{\rm c}=\infty$) results correspond to $\lambda=0$ and $\lambda=1$, respectively. The accuracy of the scaling procedure ($\lambda\rightarrow 1$) is discussed in Appendix~\ref{appD}.

\begin{figure}[t]
\includegraphics[scale=0.46]{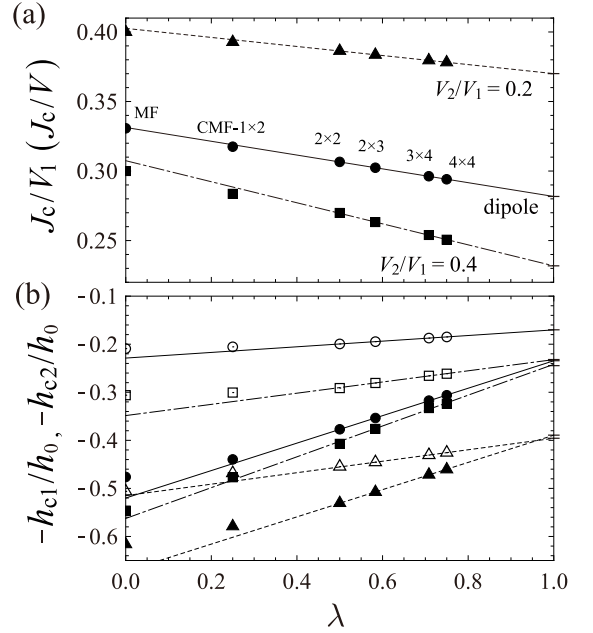}
\caption{\label{Scalings}
(a) Cluster-size scalings of the CMF data for the value of $J_{\rm c}/V_1$ ($J_{\rm c}/V$). The lines are the linear fits of the three points, $N_{\rm C}=2\times3$, $3\times4$, and $4\times4$, for $V_2/V_1=0.2$ (triangle, dashed line), $V_2/V_1=0.4$ (square, dash-dotted line), and the $V_{\rm dip}$ model (circle, solid line). (b) Cluster-size scalings of the CMF data for the phase boundaries between the CS and CSS phases, $h=\pm h_{{\rm c}1}$ (open symbols), and between the CSS and SF phases, $h=\pm h_{{\rm c}2}$ (closed symbols), at $J/J_{\rm c}=0.7$. The triangle, square, and circle symbols correspond again to the data for $V_2/V_1=0.2$, $V_2/V_1=0.4$, and the $V_{\rm dip}$ model. }
\end{figure}

Now we perform the scaling analysis to the three cases, $V_2/V_1=0.2$, $V_2/V_1=0.4$, and the $V_{\rm dip}$ model. We first consider the change in the location of the tip of the CS lobe ($J=J_{\rm c}$) with increasing $N_{\rm C}$. In all the cases, the value of $J_{\rm c}/V_1$ (or $J_{\rm c}/V$ for the $V_{\rm dip}$ model) systematically decreases with the cluster size $N_{\rm C}$, and the linear fits of the data for the three largest clusters ($N_{\rm C}=2\times3$, $3\times4$, and $4\times4$) are fairy good as shown in Fig.~\ref{Scalings}(a). 
The scaled values of $J_{\rm c}/V_1$ ($J_{\rm c}/V$) are $0.3701$, $0.2318$, and $0.2817$ for $V_2/V_1=0.2$, $V_2/V_1=0.4$, and the $V_{\rm dip}$ model, respectively. Next, we move on to the scalings of the phase boundaries between the CS, CSS, and SF phases. The extrapolations are carried out on the value of $h/h_0$ at each transition at fixed $J/J_{\rm c}$, in which $J_{\rm c}$ is the value at each cluster size. We show examples of linear fittings of the CMF data for the phase boundaries between the CS and CSS phases ($h=\pm h_{{\rm c}1}$) and between the CSS and SF phases ($h=\pm h_{{\rm c}2}$) in Fig.~\ref{Scalings}(b). We can see that in the cases of the $V_1$-$V_2$ model, the two lines of the CS-CSS and CSS-SF transitions approach each other very closely in the limit $\lambda\rightarrow 1$. Especially, the lines for $V_2/V_1=0.2$ intersect before reaching $\lambda=1$, which means that the transitions are replaced by the direct first-order transition between the CS and SF phases. On the other hand, the CSS region ($h_{{\rm c}1}<|h|<h_{{\rm c}2}$) of the $V_{\rm dip}$ model remains sufficiently large.

\begin{figure}[t]
\includegraphics[scale=0.55]{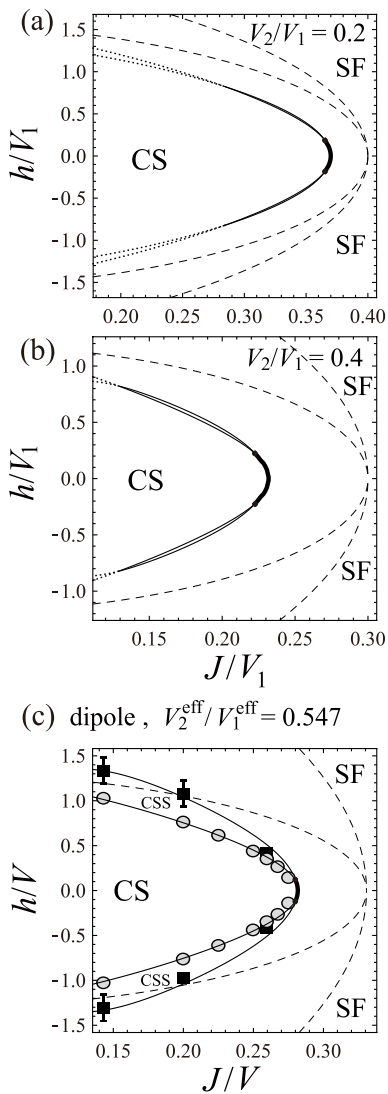}
\caption{\label{PDScaled}
The results of the scaling analyses to the phase boundaries between the CS, CSS, and SF phases for the $V_1$-$V_2$ model with (a) $V_2/V_1=0.2$ and (b) $V_2/V_1=0.4$, and for (c) the $V_{\rm dip}$ model. First-order transitions between the CS and SF phases are expected to occur at the thick solid lines and in the region between the two dotted lines, which correspond to the ``metastability limits'' of the CS and SF phases. For comparison, the MF results, Eq.~(\ref{hcV1V2}), and the QMC data (Ref.~\onlinecite{sansone-10}) are shown by the dashed lines and by the squares and circles, respectively. }
\end{figure}
Performing the same scaling analyses on the CS-CSS and CSS-SF (or CS-SF) transition boundaries for other values of $J/J_{\rm c}$, we draw the expected phase diagrams for the limit $\lambda\rightarrow 1$ in Figs.~\ref{PDScaled}(a)-\ref{PDScaled}(c). Here, the quantities on the axes are rescaled in units of $V_1$ (or $V$) by using the scaled value of $J_{\rm c}/V_1$ ($J_{\rm c}/V$), obtained in Fig.~\ref{Scalings}(a), and $h_0/V_1$ ($h_0/V$) for each case. We can see that the width of the CSS phase almost vanishes in the two cases of the $V_1$-$V_2$ model. Within the accuracy of the scaling procedure, it is difficult to provide a final conclusion on whether a very small region of the CSS phase can survive or completely disappear. However, even if the CSS region can survive, it should be too narrow to detect, and this result does not contradict the conclusion of Ref.~\onlinecite{batrouni-00}. Figure~\ref{PDScaled}(c) shows the scaled CMF result for the $V_{\rm dip}$ model, which is in surprisingly good agreement with the QMC data.~\cite{sansone-10} Unlike the two cases of the $V_1$-$V_2$ model, we can see that the CSS phase remains stable in a considerably large region of parameters.

Having obtained the above CMF results, we now summarize the difference of the $V_1$-$V_2$ and $V_{\rm dip}$ models together with the knowledge gained from the MF and LSW analyses in Sec.~\ref{3} and Sec.~\ref{4}. As for the $V_1$-$V_2$ model, we have the following dilemma: the NNN interaction $V_2$ is required to be large in order to obtain a large region of the CSS phase, according to the MF prediction in Eq.~(\ref{hcV1V2}); however, the LSW analysis showed that the larger the value of $V_2/V_1$ is, the stronger quantum fluctuations are due to the competition between the checkerboard and stripe density-wave orders. Because of this dilemma, the CSS state cannot be stabilized in a sufficiently large region of the phase diagram for both cases of small and large values of $V_2/V_1$, as shown in Figs.~\ref{PDScaled}(a) and~\ref{PDScaled}(b). On the other hand, the MF phase diagram of the $V_{\rm dip}$ model contains a rather large region of the CSS phase thanks to the large value of $V_2^{\rm eff}/V_1^{\rm eff}$ and at the same time, the long-range nature of the dipole-dipole interaction suppresses the quantum fluctuations around the MF ground state because the competition with other solid orders is weaker. For these reasons, the CSS phase can survive in the $V_{\rm dip}$ model, as shown in Fig.~\ref{PDScaled}(c), even after taking into account the effect of quantum fluctuations.

A similar discussion is applicable to the difference between triangular and kagome lattices in the hardcore Bose-Hubbard model with nearest-neighbor interaction. The MF properties of the two systems are identical except for the scale of the chemical potential, and both the systems have a large region of supersolid phase in the MF phase diagram.~\cite{murthy-97} 
However, previous QMC studies have shown that while a stable supersolid phase exists in the triangular-lattice system,~\cite{wessel-05} it has not been found in the kagome lattice.~\cite{isakov-06}
This is a similar situation to the difference between the $V_1$-$V_2$ and $V_{\rm dip}$ models. 
In this case, although the competition of different solid orders does not make a large difference between the two systems, it is known that the quantum fluctuations in the kagome lattice are much stronger than those in the triangular lattice, reflecting, e.g., the lower coordination number.~\cite{murthy-97} Because of the strong quantum fluctuations, the supersolid states in the kagome lattice are more strongly destabilized and cannot survive in the QMC calculations. Therefore, we can say that for the emergence of stable lattice supersolid states, in general, it is necessary to satisfy the following two (qualitative and quantitative) conditions: A certain long-range interaction $V_{jl}$ is required for creating solid orders, and the quantum fluctuation should be weak enough so as not to destabilize the supersolid states into phase separation. 
\section{\label{6}SUMMARY}
In conclusion, we have investigated the ground-state phase diagrams of the hardcore Bose-Hubbard model with square lattice structure and long-range interactions.
One of our main focuses is placed in understanding the role of long-range interactions in the emergence of checkerboard supersolid (CSS) states, through the comparison of the models with nearest-neighbor and next-nearest-neighbor interactions (the $V_1$-$V_2$ model) and with the dipole-dipole interaction proportional to $1/r^3$ (the $V_{\rm dip}$ model). Specifically, we discussed the reasons why the CSS states can be stable only in the case of the $V_{\rm dip}$ model, and clarified the origin of the qualitative difference between the two systems. We first showed the classical (mean-field) properties of the systems, and then discussed the strength of quantum fluctuations around them in terms of the linear spin-wave theory. Moreover, we also applied the cluster mean-field (CMF) method and its cluster-size scaling to take into account the effects of quantum fluctuations in a self-consistent way. 
We confirmed quantitative accuracy of our CMF scaling procedure~\cite{yamamoto-12} by making a comparison with the quantum Monte Carlo data in Fig.~\ref{Scalings}(c) and Appendix~\ref{appD}. In principle, this approach can be also applied to any other ordered systems including softcore bosons and higher-spin systems. Especially, our CMF method may be useful for studying frustrated systems since it is free from the minus-sign problem.

\acknowledgments
This work was supported by KAKENHI (23840054) from JSPS (D. Y.) and the Computational Materials Science Initiative (CMSI) (A.M.). The numerical calculations were partially performed on the RIKEN Integrated Cluster of Clusters (RICC) and computers at the Supercomputer Center, Institute for Solid State Physics, University of Tokyo.

\begin{appendix}
\section{three- and four-sublattice structures}\label{appA}
In the case of the $V_{\rm dip}$ model, in addition to the uniform (SF and MI) and checkerboard (CS and CSS) phases, many different types of solid and supersolid phases can appear due to the long-range nature of the dipole-dipole interaction. In general, we encounter various phases with more complex sublattice structures when the effects of the dipole-dipole force are stronger (the value of $J/V$ is smaller).~\cite{sansone-10,yamamoto-12}

\begin{figure}[b]
\includegraphics[scale=0.55]{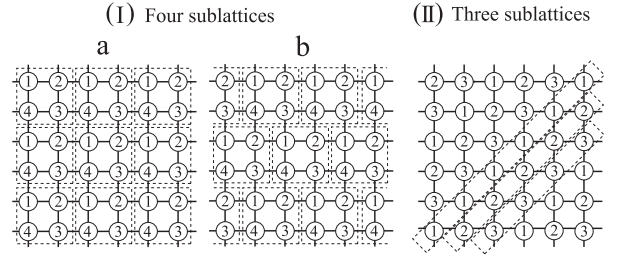}
\caption{\label{AppFig}
Schematic pictures of (I) two types of four-sublattice structures (4a and 4b) and (II) a three-sublattice structure. The lattice sites with the same number belong to the same sublattice. The dashed boxes are drawn to show clearly the sublattice structures.}
\end{figure}

To obtain the results in Sec.~\ref{3}, we restricted our MF analysis to the phases with up to the three- and four-sublattice structures shown in Figs.~\ref{AppFig}(I) and~\ref{AppFig}(II), focusing on the region of relatively large values of $J/V$ (but still less than $1$). Each state is characterized by the classical pseudospin angles $\theta_n$ ($n=1,2,3,4$ for the four sublattices; $n=1,2,3$ for the three sublattices). 
The MF energies per site for the two types (a and b) of four-sublattice phases are written as
\begin{eqnarray}
E_{0}^{(\rm 4a)}/M&=&-JS^2(\sin\theta_{\rm 1}+\sin\theta_{\rm 3})(\sin\theta_{\rm 2}+\sin\theta_{\rm 4})\nonumber\\
&&+V_1^{(\rm 4a)}S^2(\cos \theta_{\rm 1}+\cos \theta_{\rm 3})(\cos \theta_{\rm 2}+\cos \theta_{\rm 4})/2\nonumber\\
&&+V_2^{(\rm 4a)}S^2(\cos \theta_{\rm 1}\cos \theta_{\rm 3}+\cos \theta_{\rm 2}\cos \theta_{\rm 4})\nonumber\\
&&+V_3^{(\rm 4a)}S^2\sum_{n=1}^4\cos^2 \theta_{\rm n}/2\nonumber\\
&&-hS\sum_{n=1}^4\cos \theta_{\rm n}/4,\label{Ene4a}
\end{eqnarray}
where $V_1^{(\rm 4a)}\equiv 2\sum_{l_{\rm 2}} V_{j_{\rm 1}l_{\rm 2}}/z\approx 1.460V$, $V_2^{(\rm 4a)}\equiv \sum_{l_{\rm 3}} V_{j_{\rm 1}l_{\rm 3}}/z\approx 0.5162V$, and $V_3^{(\rm 4a)}\equiv \sum_{l_{\rm 1}} V_{j_{\rm 1}l_{\rm 1}}/z\approx 0.2823V$, and
\begin{widetext}
\begin{eqnarray}
E_{0}^{(\rm 4b)}/M&=&-JS^2\left[2\sin\theta_{\rm 1}\sin\theta_{\rm 2}+2\sin\theta_{\rm 3}\sin\theta_{\rm 4}+(\sin\theta_{\rm 1}+\sin\theta_{\rm 2})(\sin\theta_{\rm 3}+\sin\theta_{\rm 4})\right]/2\nonumber\\
&&+V_1^{(\rm 4b)}S^2\left[2\cos\theta_{\rm 1}\cos\theta_{\rm 2}+2\cos\theta_{\rm 3}\cos\theta_{\rm 4}+(\cos\theta_{\rm 1}+\cos\theta_{\rm 2})(\cos\theta_{\rm 3}+\cos\theta_{\rm 4})\right]/4\nonumber\\
&&+V_2^{(\rm 4b)}S^2(\cos \theta_{\rm 1}+\cos \theta_{\rm 2})(\cos \theta_{\rm 3}+\cos \theta_{\rm 4})/2\nonumber\\
&&+V_3^{(\rm 4b)}S^2\left(2\cos\theta_{\rm 1}\cos\theta_{\rm 2}+2\cos\theta_{\rm 3}\cos\theta_{\rm 4}+\sum_{n=1}^4\cos^2 \theta_{\rm n}\right)/4\nonumber\\
&&-hS\sum_{n=1}^4\cos \theta_{\rm n}/4,\label{Ene4b}
\end{eqnarray}
\end{widetext}
where $V_1^{(\rm 4b)}\equiv 2\sum_{l_{\rm 1}} (V_{j_{\rm 2}l_{\rm 1}}-V_{j_{\rm 1}l_{\rm 1}})/z\approx 0.9077V$, $V_2^{(\rm 4b)}\equiv \sum_{l_{\rm 1}} (2V_{j_{\rm 3}l_{\rm 1}}+V_{j_{\rm 1}l_{\rm 1}}-V_{j_{\rm 2}l_{\rm 1}})/z\approx 0.7923V$, and $V_3^{(\rm 4b)}\equiv 2\sum_{l_{\rm 1}} V_{j_{\rm 1}l_{\rm 1}}/z\approx 0.5584V$. 
The above expressions are formally identical to those of the hardcore Bose-Hubbard model with up to the third-nearest-neighbor interactions ($V_1$, $V_2$, and $V_3$); we have only to replace $V_n$ ($n=1,2,$ and $3$) with the effective interactions $V_n^{(\rm 4a)}$ for the a-type structure or $V_n^{(\rm 4b)}$  for the b-type structure in the corresponding expressions of the MF energies. 
On the other hand, the MF energy for the three-sublattice structure of Fig.~\ref{AppFig}(c) is given by
\begin{widetext}
\begin{eqnarray}
E_{0}^{(\rm 3)}/M&=&-4JS^2(\sin\theta_{\rm 1}\sin\theta_{\rm 2}+\sin\theta_{\rm 2}\sin\theta_{\rm 3}+\sin\theta_{\rm 3}\sin\theta_{\rm 1})/3+2V_1^{(\rm 3)}S^2(\cos\theta_{\rm 1}\cos\theta_{\rm 2}+\cos\theta_{\rm 2}\cos\theta_{\rm 3}+\cos\theta_{\rm 3}\cos\theta_{\rm 1})/3\nonumber\\
&&+V_2^{(\rm 3)}S^2\left(\cos\theta_{\rm 1}\cos\theta_{\rm 2}+\cos\theta_{\rm 2}\cos\theta_{\rm 3}+\cos\theta_{\rm 3}\cos\theta_{\rm 1}+\sum_{n=1}^3\cos^2 \theta_{\rm n}\right)/3-hS\sum_{n=1}^3\cos \theta_{\rm n}/3,\label{Ene3}
\end{eqnarray}
\end{widetext}
where $V_1^{(\rm 3)}\equiv \sum_{l_{\rm 1}} (2V_{j_{\rm 2}l_{\rm 1}}-V_{j_{\rm 1}l_{\rm 1}})/z\approx 1.317V$ and $V_2^{(\rm 3)}\equiv 2\sum_{l_{\rm 1}} V_{j_{\rm 1}l_{\rm 1}}/z\approx 0.9417V$.

To obtain the ground-state phase diagram in Fig.~\ref{MFPDDip}, we carried out the minimization of the MF energies in the standard way. First, we minimized separately the MF energies given in Eqs.~(\ref{Ene4a}), (\ref{Ene4b}), and (\ref{Ene3}) with respect to the angles $\theta_n$, and then compared the three minimized values of the MF energies. For example, if such a minimization scheme leads to a solution of the a-type four-sublattice structure with $\theta_2=\theta_4$, it means that the ground state is in the SS2a phase within the MF approximation. As for the other phases seen in Fig.~\ref{MFPDDip}, the assumption of the b-type four-sublattice structure includes solutions of the SS2b and $\rho=1/4$ solid states and the three sublattice structure given in Fig.~\ref{AppFig}(II) includes the $\rho=1/3$ solid state. 

\section{Details of the LSW analysis}\label{appB} 
We present here the details of the LSW calculations. 
In the rotated frame of Eq.~(\ref{rotation}), the pseudospin Hamiltonian [Eq.~(\ref{spinH})] is rewritten as 
\begin{eqnarray}
\hat{H}_{\rm spin}&=&
-\frac{1}{2}\sum_{j,l}\sum_{\mu,\nu=x,y,z}
\left(\tilde{S}_j^{\mu} I_{jl}^{\mu\nu}\tilde{S}_l^\nu\right)\nonumber\\
&&- \sum_j h_j^x \tilde{S}^{x}_j- \sum_j h_j^z \tilde{S}^{z}_j,
\label{eq:spinH2}
\end{eqnarray}
where
\begin{eqnarray}
\begin{array}{ccl}
I_{jl}^{xx}&=&2J_{jl}\cos \theta_j\cos \theta_l-V_{jl}\sin \theta_j\sin \theta_l,~~
I_{jl}^{yy}=2J_{jl},\\
I_{jl}^{zz}&=&2J_{jl}\sin \theta_j\sin \theta_l-V_{jl}\cos \theta_j\cos \theta_l,\\
I_{jl}^{zx}&=&I_{jl}^{xz}=2J_{jl}\sin \theta_j\cos \theta_l+V_{ij}\cos \theta_j\sin \theta_l,\\
I_{jl}^{xy}&=&I_{jl}^{yx}=I_{jl}^{yz}=I_{jl}^{zy}=0,\\
h_j^x&=&-h\sin \theta_j,~h_j^z=h\cos \theta_j.
\end{array}
\end{eqnarray}
After performing the HP transformation in Eq.~(\ref{HPtr}), the quadratic part of the pseudospin Hamiltonian is obtained as
\begin{eqnarray}
\hat{H}_{2}&=&-\frac{S}{2}\sum_{j,l}\Bigg[\left(I_{jl}^{xx}+I_{jl}^{yy}\right)\hat{b}_j^\dagger\hat{b}_l\Bigg]\nonumber\\
&&-\frac{S}{4}\sum_{j,l}\Bigg[\left(I_{jl}^{xx}-I_{jl}^{yy}\right)\hat{b}_j\hat{b}_l+{\rm H.c.}\Bigg]\nonumber\\&&+\sum_j\left[\left(h_j^z+S\sum_l I_{jl}^{zz}\right)\hat{b}_j^\dagger\hat{b}_j\right]. 
\end{eqnarray}
In order to diagonalize $\hat{H}_{2}$, the use of the equation of motion method with the Green's functions may be more convenient than the usual Bogoliubov transformation technique for the states with complex sublattice structures. 

First, we define the retarded commutator Green's function in a matrix form as
\begin{eqnarray}
\langle\langle \hat{\bf b}_j(t);\hat{\bf b}^\dagger_l(t')\rangle\rangle= -i\theta(t-t')\langle [\hat{\bf b}_j(t),\hat{\bf b}^\dagger_l(t')] \rangle
\end{eqnarray}
with
\begin{eqnarray}
\hat{\bf b}_j\equiv \left(\begin{array}{c}
\hat{b}_j\\
\hat{b}_j^\dagger
\end{array}\right)~{\rm and}~\theta(t-t')=\left\{ \begin{array}{ll}
1 &  (t>t') \\
0 & (t<t')
\end{array} \right..
\end{eqnarray}
The Green's function satisfies the following equation of motion: 
\begin{eqnarray}
i\frac{\partial}{\partial t}\langle\langle \hat{\bf b}_j(t);\hat{\bf b}^\dagger_l(t')\rangle\rangle&=&\delta(t-t')\langle[\hat{\bf b}_j(t),\hat{\bf b}^\dagger_l(t')]\rangle\nonumber\\&&+\langle\langle [\hat{\bf b}_j,\hat{H}_{2}](t);\hat{\bf b}^\dagger_l(t')\rangle\rangle.\label{EqofMotion}
\end{eqnarray}
To solve the above equation, we perform Fourier transform of the Green's function ${\bf G}_{jl}(t-t')\equiv \langle\langle \hat{\bf b}_j(t);\hat{\bf b}^\dagger_l(t')\rangle\rangle$ into energy space
\begin{eqnarray}
{\bf G}_{jl}(t-t')=\int_{-\infty}^{\infty}\frac{d \omega}{2\pi}{\bf G}_{jl}(\omega)e^{-i \omega (t-t')}
\end{eqnarray}
and into momentum space. It is to be noted that the number of sites in the ``unit cell'' varies depending on the sublattice structure. 
For example, since the lattice sites are divided into two square sublattices [A and B in Fig.~\ref{sublattice}(I)] for the CS and CSS states, the Fourier transformation into the momentum space should be performed on each sublattice: 
\begin{eqnarray}
{\bf G}_{j_{\alpha}l_{\beta}}=\frac{2}{N}\sum_{\bf q}'{\bf G}^{\alpha\beta}_{\bf q}e^{-i{\bf q}\cdot ({\bf r}_{j_{\alpha}}-{\bf r}_{{l_{\beta}}})},
\end{eqnarray}
where the subscripts $\alpha$, $\beta$ denote the sublattice index (A or B), and the sum is taken over the $N/2$ ${\bf q}$-values in the reduced Brillouin zone. Now, we can rewrite the equation of motion for the Green's functions in a $4\times 4$ matrix form:
\begin{widetext}
\begin{eqnarray}
\omega \left(\begin{array}{cc}
{\bf G}_{\bf q}^{\rm AA}&{\bf G}_{\bf q}^{\rm AB}\\
{\bf G}_{\bf q}^{\rm BA}&{\bf G}_{\bf q}^{\rm BB}
\end{array}\right)&=&\left(\begin{array}{cc}
\mbox{\boldmath $\sigma$}_z&{\bf 0}\\
{\bf 0}&\mbox{\boldmath $\sigma$}_z
\end{array}\right)+\left(\begin{array}{cc}
\mbox{\boldmath $\Gamma$}_{\bf q}^{\rm AA}&\mbox{\boldmath $\Gamma$}_{\bf q}^{\rm AB}\\
\mbox{\boldmath $\Gamma$}_{\bf q}^{\rm BA}&\mbox{\boldmath $\Gamma$}_{\bf q}^{\rm BB}
\end{array}\right)\left(\begin{array}{cc}
{\bf G}_{\bf q}^{\rm AA}&{\bf G}_{\bf q}^{\rm AB}\\
{\bf G}_{\bf q}^{\rm BA}&{\bf G}_{\bf q}^{\rm BB}
\end{array}\right).\label{EqofMotionq}
\end{eqnarray}
The $2\times 2$ submatrices in Eq.~(\ref{EqofMotionq}) are given by
\begin{eqnarray}
\mbox{\boldmath $\sigma$}_z=\left(\begin{array}{cc}
1&0\\
0&-1
\end{array}\right)~~{\rm and}~~\mbox{\boldmath $\Gamma$}_{\bf q}^{\alpha\beta}=\left(\begin{array}{cc}
\Gamma_{{\bf q},11}^{\alpha\beta}&\Gamma_{{\bf q},12}^{\alpha\beta}\\
-\Gamma_{{\bf q},12}^{\alpha\beta}&-\Gamma_{{\bf q},11}^{\alpha\beta}
\end{array}\right)~~(\alpha,\beta={\rm A},{\rm B}),
\end{eqnarray}
where
\begin{eqnarray}
\Gamma_{{\bf q},11}^{\rm AA}&=&H\cos \theta_{\rm A}+4S(2J\sin \theta_{\rm A}\sin \theta_{\rm B}-V_1\cos \theta_{\rm A}\cos \theta_{\rm B}-V_2\cos^2 \theta_{\rm A})+2S\sin^2 \theta_{\rm A}V^{(2)}_{\bf q},\nonumber\\
\Gamma_{{\bf q},11}^{\rm BB}&=&H\cos \theta_{\rm B}+4S(2J\sin \theta_{\rm A}\sin \theta_{\rm B}-V_1\cos \theta_{\rm A}\cos \theta_{\rm B}-V_2\cos^2 \theta_{\rm B})+2S\sin^2 \theta_{\rm B}V^{(2)}_{\bf q},\nonumber\\
\Gamma_{{\bf q},12}^{\rm AA}&=&2S\sin^2 \theta_{\rm A}V^{(2)}_{\bf q},~~~\Gamma_{{\bf q},12}^{\rm BB}=2S\sin^2 \theta_{\rm B}V^{(2)}_{\bf q},\nonumber\\
\Gamma_{{\bf q},11}^{\rm AB}&=&\Gamma_{{\bf q},11}^{\rm BA}=-4JS(\cos \theta_{\rm A}\cos \theta_{\rm B}+1)\gamma_{\bf q}+2S\sin \theta_{\rm A}\sin \theta_{\rm B}V^{(1)}_{\bf q},\nonumber\\
\Gamma_{{\bf q},12}^{\rm AB}&=&\Gamma_{{\bf q},12}^{\rm BA}=-4JS(\cos \theta_{\rm A}\cos \theta_{\rm B}-1)\gamma_{\bf q}+2S\sin \theta_{\rm A}\sin \theta_{\rm B}V^{(1)}_{\bf q}.
\end{eqnarray}
\end{widetext}
Here, $V_1$ and $V_2$ should be replaced with $V_1^{\rm eff}$ and $V_2^{\rm eff}$, respectively, for the $V_{\rm dip}$ model. The Fourier factor $\gamma_{\bf q}$ is given by $\gamma_{\bf q}=[\cos (q_x d)+\cos (q_y d)]/2$, and $V^{(1,2)}_{\bf q}$ are defined by 
\begin{subequations}
\begin{eqnarray}
V^{(1)}_{\bf q}&\equiv& \frac{1}{z}\sum_{l_{\rm B}} V_{j_{\rm A}l_{\rm B}}e^{i{\bf q}\cdot({\bf r}_{j_{\rm A}}-{\bf r}_{l_{\rm B}})},\\
V^{(2)}_{\bf q}&\equiv& \frac{1}{z}\sum_{l_{\rm A}} V_{j_{\rm A}l_{\rm A}}e^{i{\bf q}\cdot({\bf r}_{j_{\rm A}}-{\bf r}_{l_{\rm A}})}. 
\end{eqnarray}\label{Vq}\end{subequations}
For the $V_1$-$V_2$ model, these can be simply written as $V^{(1)}_{\bf q}=V_1\gamma_{\bf q}$ and $V^{(2)}_{\bf q}=V_2\cos (q_x d)\cos (q_y d)$.

We can obtain the spin-wave excitation spectra $\omega ({\bf q})$, which correspond to the poles of the Green's functions, by solving the equation
\begin{eqnarray}
{\rm det}\left[\omega ({\bf q}) \hat{\bf 1}-\left(\begin{array}{cc}
\mbox{\boldmath $\Gamma$}_{\bf q}^{\rm AA}&\mbox{\boldmath $\Gamma$}_{\bf q}^{\rm AB}\\
\mbox{\boldmath $\Gamma$}_{\bf q}^{\rm BA}&\mbox{\boldmath $\Gamma$}_{\bf q}^{\rm BB}
\end{array}\right)\right]=0.
\end{eqnarray}
Moreover, we can calculate $\langle \hat{b}_j^\dagger \hat{b}_j\rangle$ by applying the spectral theorem to the Green's functions 
\begin{eqnarray}
&&\left(\begin{array}{cc}
{\bf G}_{\bf q}^{\rm AA}&{\bf G}_{\bf q}^{\rm AB}\\
{\bf G}_{\bf q}^{\rm BA}&{\bf G}_{\bf q}^{\rm BB}
\end{array}\right)\nonumber\\
&&=\left[\omega \hat{\bf 1}-\left(\begin{array}{cc}
\mbox{\boldmath $\Gamma$}_{\bf q}^{\rm AA}&\mbox{\boldmath $\Gamma$}_{\bf q}^{\rm AB}\\
\mbox{\boldmath $\Gamma$}_{\bf q}^{\rm BA}&\mbox{\boldmath $\Gamma$}_{\bf q}^{\rm BB}
\end{array}\right)\right]^{-1}\!\!\left(\begin{array}{cc}
\mbox{\boldmath $\sigma$}_z&{\bf 0}\\
{\bf 0}&\mbox{\boldmath $\sigma$}_z
\end{array}\right).\label{GF}
\end{eqnarray}
The number of spin waves on each sublattice ($\alpha=$A or B) is given by
\begin{eqnarray}
\langle \hat{b}_{j_\alpha}^\dagger \hat{b}_{j_\alpha}\rangle&=&\frac{i}{2\pi}\frac{2}{N}\sum_{\bf q}^\prime\lim_{\delta\to 0}\int_{-\infty}^{\infty} \frac{d\omega}{e^{\beta\omega}-1}\nonumber\\
&&\times \left[ G_{{\bf q},11}^{\alpha\alpha}(\omega+i\delta)-G_{{\bf q},11}^{\alpha\alpha}(\omega-i\delta)\right],
\end{eqnarray}
where $G_{{\bf q},11}^{\alpha\alpha}$ is the $(1,1)$-component of ${\bf G}_{{\bf q}}^{\alpha\alpha}$ in Eq.~(\ref{GF}).

We took here the case of the checkerboard phases as an example. The extension to other sublattice structures is straightforward. 
For instance, the calculation for the SS2a phase, in which the lattice sites are divided into four square sublattices with lattice constant $2d$, requires the use of a $8\times8$ matrix form instead of Eq.~(\ref{EqofMotionq}). 

\section{Details of the CMF calculations}\label{appC}
\begin{figure}[b]
\includegraphics[scale=0.6]{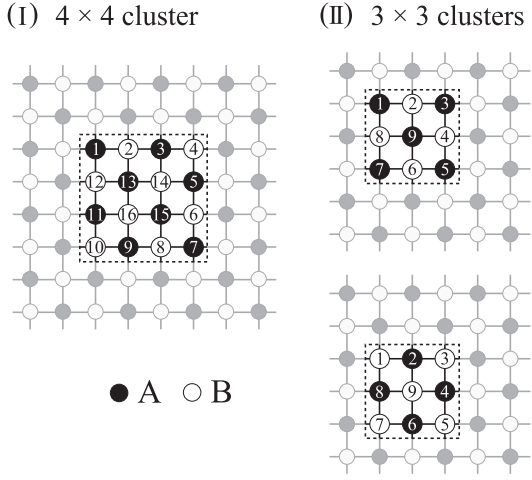}
\caption{\label{ExEmbed2}
The same as in Figs.~\ref{ExEmbed}(I) and~\ref{ExEmbed}(II) with the site labels ($1,2,3,\cdots$) within the clusters. }
\end{figure}

We present here the explicit forms of the effective fields $h_j^{z,{\rm eff}}$ and $h_j^{x,{\rm eff}}$ in Eq.~(\ref{spinH_C}) for the reader's convenience. Here, as examples, we show the expressions in the CMF calculations for checkerboard phases with $N_{\rm C}=3\times 3$ and $4\times 4$ clusters [see Figs.~\ref{ExEmbed2}(I) and~\ref{ExEmbed2}(II)]. 

In the case of $N_{\rm C}=4\times 4$, we have only one choice, Fig.~\ref{ExEmbed2}(I), for the cluster embedded into the background sublattice structure. In that sense, we can say that the use of clusters with even numbers of sites is compatible with the two-sublattice checkerboard pattern of solid orders. The explicit forms of the effective fields for site ``1'' in Fig.~\ref{ExEmbed2}(I) were already shown in Eqs.~(\ref{hzx1V1V2}) and~(\ref{hzx1Vdip}). For the other sites, we have
\begin{widetext}
\begin{eqnarray}\begin{array}{rcl}
h_7^{z,{\rm eff}}&=&h_1^{z,{\rm eff}},~h_{4,10}^{z,{\rm eff}}=-2V_1 m^{z}_{\rm A}-3V_2 m^{z}_{\rm B},~h_{3,5,9,11}^{z,{\rm eff}}= -V_1 m^{z}_{\rm B}-2V_2 m^{z}_{\rm A},~h_{2,6,8,12}^{z,{\rm eff}}=-V_1 m^{z}_{\rm A}-2V_2 m^{z}_{\rm B},\\
h_{13,14,15,16}^{z,{\rm eff}}&=&0,~
h_7^{x,{\rm eff}}=h_1^{x,{\rm eff}},~h_{4,10}^{x,{\rm eff}}=4J m^{x}_{\rm A},~h_{3,5,9,11}^{x,{\rm eff}}= 2J m^{x}_{\rm B},~
h_{2,6,8,12}^{x,{\rm eff}}= 2J m^{x}_{\rm A},~
h_{13,14,15,16}^{x,{\rm eff}}=0
\end{array}\label{hzx16V1V2}\end{eqnarray}
for the $V_1$-$V_2$ model and 
\begin{eqnarray}\begin{array}{rcl}
h_7^{z,{\rm eff}}&=& h_1^{z,{\rm eff}},~\\
h_{4,10}^{z,{\rm eff}}&=&\displaystyle -\left(4V_1^{\rm eff}-2V-\frac{2V}{\sqrt{5}^3}-\frac{2V}{3^3}-\frac{2V}{\sqrt{13}^3}\right) m^{z}_{\rm A}-\left(4V_2^{\rm eff}-\frac{V}{\sqrt{2}^3}-\frac{2V}{2^3}-\frac{V}{\sqrt{8}^3}-\frac{2V}{\sqrt{10}^3}-\frac{V}{\sqrt{18}^3}\right) m^{z}_{\rm B},\\
h_{3,5,9,11}^{z,{\rm eff}}&=&\displaystyle -\left(4V_1^{\rm eff}-3V-\frac{3V}{\sqrt{5}^3}-\frac{V}{3^3}-\frac{V}{\sqrt{13}^3}\right) m^{z}_{\rm B}-\left(4V_2^{\rm eff}-\frac{2V}{\sqrt{2}^3}-\frac{2V}{2^3}-\frac{V}{\sqrt{8}^3}-\frac{2V}{\sqrt{10}^3}\right) m^{z}_{\rm A},\\
h_{2,6,8,12}^{z,{\rm eff}}&=&\displaystyle -\left(4V_1^{\rm eff}-3V-\frac{3V}{\sqrt{5}^3}-\frac{V}{3^3}-\frac{V}{\sqrt{13}^3}\right) m^{z}_{\rm A}-\left(4V_2^{\rm eff}-\frac{2V}{\sqrt{2}^3}-\frac{2V}{2^3}-\frac{V}{\sqrt{8}^3}-\frac{2V}{\sqrt{10}^3}\right) m^{z}_{\rm B},\\
h_{13,15}^{z,{\rm eff}}&=&\displaystyle -\left(4V_1^{\rm eff}-4V-\frac{4V}{\sqrt{5}^3}\right) m^{z}_{\rm B}-\left(4V_2^{\rm eff}-\frac{4V}{\sqrt{2}^3}-\frac{2V}{2^3}-\frac{V}{\sqrt{8}^3}\right) m^{z}_{\rm A},\\
h_{14,16}^{z,{\rm eff}}&=&\displaystyle -\left(4V_1^{\rm eff}-4V-\frac{4V}{\sqrt{5}^3}\right) m^{z}_{\rm A}-\left(4V_2^{\rm eff}-\frac{4V}{\sqrt{2}^3}-\frac{2V}{2^3}-\frac{V}{\sqrt{8}^3}\right) m^{z}_{\rm B},\\
h_7^{x,{\rm eff}}&=&h_1^{x,{\rm eff}},~h_{4,10}^{x,{\rm eff}}=4J m^{x}_{\rm A},~h_{3,5,9,11}^{x,{\rm eff}}= 2J m^{x}_{\rm B},~
h_{2,6,8,12}^{x,{\rm eff}}= 2J m^{x}_{\rm A},~
h_{13,14,15,16}^{x,{\rm eff}}=0
\end{array}\label{hzx16Vdip}\end{eqnarray}
for the $V_{\rm dip}$ model. Using these expressions, we solved the CMF self-consistent equations, Eqs.~(\ref{SCEq}), for the mean fields $m^{z,x}_{\rm A}$ and $m^{z,x}_{\rm B}$.

Next, we show the case of $N_{\rm C}=3\times 3$, although we did not use this size of cluster for the scaling analysis in this paper. In this case, we have to treat the two clusters given in Fig.~\ref{ExEmbed2}(II) and the two corresponding cluster Hamiltonians $\hat{H}_{C_1}$ and $\hat{H}_{C_2}$. 
The effective fields for the upper cluster in Fig.~\ref{ExEmbed2}(II) are given by
\begin{eqnarray}\begin{array}{rcl}
h_{1,3,5,7}^{z,{\rm eff}}&=&-2V_1 m^{z}_{\rm B}-3V_2 m^{z}_{\rm A},~h_{2,4,6,8}^{z,{\rm eff}}= -V_1 m^{z}_{\rm A}-2V_2 m^{z}_{\rm B},~h_{9}^{z,{\rm eff}}=0,\\
h_{1,3,5,7}^{x,{\rm eff}}&=&4J m^{x}_{\rm B},~h_{2,4,6,8}^{x,{\rm eff}}= 2J m^{x}_{\rm A},~h_{9}^{x,{\rm eff}}=0
\end{array}\label{hzx9V1V2}\end{eqnarray}
for the $V_1$-$V_2$ model and 
\begin{eqnarray}\begin{array}{rcl}
h_{1,3,5,7}^{z,{\rm eff}}&=&\displaystyle -\left(4V_1^{\rm eff}-2V-\frac{2V}{\sqrt{5}^3}\right) m^{z}_{\rm B}-\left(4V_2^{\rm eff}-\frac{V}{\sqrt{2}^3}-\frac{2V}{2^3}-\frac{V}{\sqrt{8}^3}\right) m^{z}_{\rm A},\\
h_{2,4,6,8}^{z,{\rm eff}}&=& \displaystyle -\left(4V_1^{\rm eff}-3V-\frac{2V}{\sqrt{5}^3}\right) m^{z}_{\rm A}-\left(4V_2^{\rm eff}-\frac{2V}{\sqrt{2}^3}-\frac{V}{2^3}\right) m^{z}_{\rm B},\\
h_{9}^{z,{\rm eff}}&=&\displaystyle -\left(4V_1^{\rm eff}-4V\right) m^{z}_{\rm B}-\left(4V_2^{\rm eff}-\frac{4V}{\sqrt{2}^3}\right) m^{z}_{\rm A},\\
h_{1,3,5,7}^{x,{\rm eff}}&=&4J m^{x}_{\rm B},~h_{2,4,6,8}^{x,{\rm eff}}= 2J m^{x}_{\rm A},~h_{9}^{x,{\rm eff}}=0
\end{array}\label{hzx9Vdip}\end{eqnarray}
\end{widetext}
for the $V_{\rm dip}$ model. The expressions for the lower cluster in Fig.~\ref{ExEmbed2}(II) can be obtained by exchanging $m^{z,x}_{\rm A}$ and $m^{z,x}_{\rm B}$ in Eqs.~(\ref{hzx9V1V2}) and (\ref{hzx9Vdip}).

\section{Accuracy estimation of the CMF method}\label{appD}
To evaluate the quantitative accuracy of our CMF scaling procedure, we compare the results for the simple model only with the NN interaction (i.e., the $V_1$-$V_2$ model with $V_2=0$) with the corresponding QMC data. It is known that no CSS phase appears for $V_2=0$ and the system exhibits only the direct first-order transition from the CS to SF phase.~\cite{batrouni-00,kimura-11} Moreover, since the quantum fluctuations do not lift the degeneracy at $(J,h)=(0.5V_1,0)$ (called the Heisenberg point), the value of $J_{\rm c}=0.5V_1$ does not change with the cluster size $N_{\rm C}$. 
\begin{figure}[t]
\includegraphics[scale=0.46]{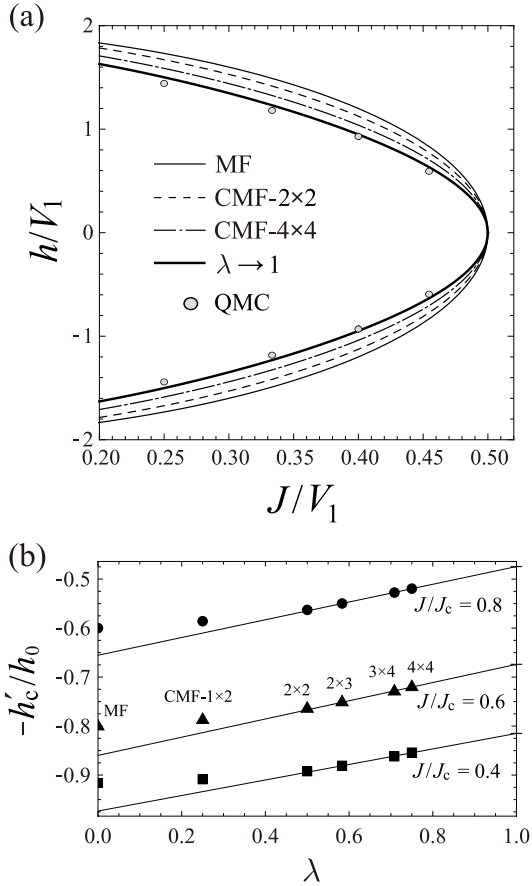}
\caption{\label{V1PDScaling}
(a) The CS-SF phase boundaries for the NN interaction model by the MF (thin solid line) and CMF methods with $N_{\rm C}=2\times 2$ (dashed line) and $4\times 4$ (dash-dotted line). The thick solid line indicates the scaled value of the CMF data with the three largest clusters of $N_{\rm C}=2\times 3$, $3\times 4$, and $4\times 4$. For comparison, we also plot the numerical data obtained by using the QMC method based on the directed loop algorithm (Ref.~\onlinecite{syljuasen-02}) for $16\times 16$ lattice sites (circles). The value $J_{\rm c}=0.5V_1$ is constant with respect to $N_{\rm C}$ as well as $h_0=2V_1$. (b) Cluster-size scalings ($\lambda\rightarrow 1$) of the CMF data for the values of $h/h_0$ at the CS-SF first-order transition of the NN interaction model. We plot the values on the lower branch of the transition lines $h=\pm h_{\rm c}^\prime$ at $J/J_{\rm c}=0.8$, $0.6$, and $0.4$. The solid lines are the linear fits of the three samples ($N_{\rm C} = 2\times 3$, $3\times4$, and $4\times 4$). }
\end{figure}

In Fig.~\ref{V1PDScaling}(a), we see that the region of the CS phase gradually gets more narrow with increasing the size of the cluster used in the CMF calculations, which means that the CS phase is first overestimated in the MF theory and then gradually improved by taking into account correlation effects within the cluster. This kind of systematic behavior is also seen in other types of cluster extensions of the MF theory.~\cite{mcintosh-12} 
We perform a linear extrapolation toward $\lambda=1$ using the three samples of $N_{\rm C}=2\times 3$, $3\times4$, and $4\times 4$, which is indicated as the thick solid curve in Fig.~\ref{V1PDScaling}(a). We can see that the scaled value and the QMC data are in good accordance, although the extrapolation still slightly overestimates the CS phase. The linear fits of the three points are quite good as shown in Fig.~\ref{V1PDScaling}(b). However, it gets worse for very small values of $J/V_1$, which is attributed to the fact that the cluster-``shape'' dependence becomes more pronounced. The scaled values should be improved by using the results of larger-size clusters as sample data for the extrapolation. 
\end{appendix}

\end{document}